\documentclass{aa}
\usepackage{natbib}

\bibpunct{(}{)}{;}{a}{}{,} 

\usepackage{mathptmx}
\usepackage[T1]{fontenc}
\usepackage[varg]{txfonts}
\usepackage{babel}
\usepackage{array}
\usepackage{textcomp}
\usepackage{multirow}
\usepackage{amsmath}
\usepackage{amssymb}
\usepackage{amsfonts}
\usepackage{lscape}

\usepackage{adjustbox}

\usepackage{longtable}
\usepackage{rotating}
\usepackage{psfrag,{graphicx}}

\usepackage{graphicx}
\usepackage{txfonts}
\usepackage{amsmath}

\makeatother

\begin{document}
\newcommand{\Msun}{$M_\odot~$}
\newcommand{\Rsun}{$R_\odot~$}
\renewcommand{\deg}{$^{\circ}$}
\newcommand{\Ro}{$R_{\odot}~$}
\newcommand{\Rgal}{$R_{\mathrm{gal}}~$}

   \title{Evolution over time of the Milky Way's disc shape}

   \author{E.~B. Am\^{o}res
          \inst{1,2}
          \and
          A.~C. Robin  \inst{2}
          \and
          C.~Reyl\'e  \inst{2}
          }
       \institute{UEFS, Departamento de F\'{i}sica, Av. Transnordestina, S/N, Novo Horizonte, Feira de Santana, CEP
         44036-900, BA, Brazil. \\
         \email{ebamores@uefs.br}
         \and
         Institut UTINAM, CNRS UMR 6213, Observatoire des Sciences de l'Univers THETA Franche-Comt\'{e} Bourgogne,
         Univ. Bourgogne Franche-Comt\'{e}, 41 bis avenue de l'Observatoire, 25000,
         Besan\c{c}on, France. \\
             }
   \date{Received: March 8, 2016 ; accepted }

  \abstract
   {Galactic structure studies
   can be used as a path to constrain the scenario of formation and evolution of our Galaxy. The dependence with the age of stellar population parameters would be linked with the history of star formation and dynamical evolution.}
   {We aim to investigate the structures of the outer Galaxy, such as the scale length, disc truncation, warp and flare  of the thin disc and study their
   dependence with age by using 2MASS data and a population synthesis model (the so-called Besan\c{c}on Galaxy Model).}
   {We have used a genetic algorithm to adjust the parameters on the observed colour-magnitude diagrams at longitudes $80^\circ \le \ell \le 280^\circ$ for $|b| \le 5.5^\circ$. We explored parameter degeneracies and uncertainties.}
   {We identify a clear dependence of the thin disc scale length, warp and flare shapes with age. The scale length is found to vary between 3.8 kpc for the youngest to
   about 2 kpc for the oldest. The warp shows a complex structure, clearly asymmetrical with a node angle changing with age
   from approximately 165$^\circ$ for old stars to 195$^\circ$ for
   young stars. The outer disc is also flaring with a scale height that varies by a factor of two between the solar neighbourhood
   and a Galactocentric distance of 12 kpc.}
   {We conclude that the thin disc scale length is in good agreement with the inside-out formation scenario and that the outer disc is
   not in dynamical equilibrium. The warp deformation with time may provide some clues to its origin.}

   \keywords{Galaxy: evolution --
             Galaxy: formation --
             Galaxy: fundamental parameters --
             Galaxy: general --
             Galaxy: stellar content --
             Galaxy: structure.
               }
   \maketitle
%
\section{Introduction}

Stellar ages are difficult to measure. They can be studied only in
rare cases when accurate, high-resolution spectroscopy, or
asteroseismology is available and for limited evolution stages
during which the stars quickly change either their luminosity or
colours. Alternatively, ages are often indirectly deduced using
chemical or kinematical criteria, in which case their
determinations are more model dependent. On the other hand,
scenarios of Galactic formation and evolution are best
investigated when ages are available.

For this reason, Galactic structure parameters, such as scale
length, scale heights, warp and flare have mainly been determined
for the Milky Way without accounting for time dependence. In a few
exceptions, tracers of different ages have been used to measure
these structures. This could partly explain why the thin disc
scale length in the literature has been claimed to have values in
between 1 to 5 kpc. However, it is noticeable that since the year
2000, most results tend towards short values, smaller than 2.6
kpc: \cite{Chen2001}: 2.25 kpc, \cite{Siegel2002}: 2-2.5 kpc,
\cite{LC2002}: 1.97 $^{+0.15}_{0.12}$ kpc,
\cite{Cabrera-Lavers2005}: 2.1 kpc, \cite{Bilir2006}: 1.9 kpc,
\cite{Karaali2007}: 1.65-2.52 kpc, \cite{Juric2008}: 2.6 $\pm$ 0.5
kpc, \cite{Yaz2010}: 1.-1.68 kpc, \cite{Robin2012}: 2.5 kpc. A few
studies have given large values, but mostly with large error bars
and correlated parameters: \cite{Larsen2003}: 3.5 kpc,
\cite{Chang2011}: 3.7 $\pm$ 1.0 kpc, \cite{McMillan2011}: 2.90
$\pm$ 0.22 kpc, \cite{Cheng2012}: 3.4$^{+2.8}_{-0.9}$ kpc,
\cite{bovy2012}: 3.5 $\pm$ 0.2 kpc but changing from 2.4 to 4.4
kpc with metallicity, and \cite{Cignoni2008}: 2.24-3.00 kpc from
open clusters.

To solve this open debate, it is worth investigating further
whether these different studies are considering the same
populations, with the same age, and what is the accuracy of their
distance estimates and the possible biases. It can be more
efficient to consider studies, whenever possible, in which the
tracers are better identified, and their ages are more or less
known or estimated.

The HI disc is known to have a long scale length
\citep{Kalberla2009}. Using the Leiden/Argentine/Bonn (LAB) survey
\citep{Kalberla2005}, \cite{Kalberla2008} have analysed the global
properties of the HI distribution in the outer Galaxy, determining
its mean surface densities, rotation curve, and mass distribution.
They obtained a radial exponential scale length of 3.75 kpc in the
mid-plane in the Galactocentric distance range 7-35 kpc.

It has been shown that the young object density laws (OB, A stars,
Cepheids, open clusters, etc.) follow longer scale lengths than
the mean thin disc. \cite{Sale2010} used A type stars to determine
the scale length in the outer Galaxy and showed that those stars
of mean age 100 Myr have a typical scale length of 3 kpc.

The density profiles and their dependence with age are crucial to
understand the scenario of Galactic evolution and formation. If
the Galaxy was formed by a process of inside-out formation as
proposed by many authors \citep{Larson1976, Sommer2003,
Rahimi2011, Brook2012, Haywood2013}, among others, one can expect
the scale length to be time dependent. More precisely, the young
disc scale length should be larger than the old disc scale length.
However, radial migration in the disc can also perturb this simple
idea.

\cite{bovy2012} noticed that the mean scale length of high
metallicity thin disc stars (probably younger in the mean) is
significantly shorter than the one of lower metallicity stars
(probably older). At the first sight, this can be contradictory to
the scenario of inside-out formation. However, thin disc
metal-poor stars are also more typical of the outer disc and could
reach the solar neighbourhood by migration and high metallicity
stars are the Sun position can be older if they come from the
inner Galaxy. \cite{Haywood2013} suggested that the evolution of
the thin outer disc is disconnected from the thin inner disc and
the thin disc scale length would vary with time. It is still a
question whether the structure of the outer disc is in a steady
state or perturbed by active merging that could manifest by the
recently discovered sub-structures like the Monoceros ring
\citep{Newberg2002, Rocha-Pinto2003} or the Canis Major
overdensity \citep{Martin2004}.

It is well known that like many large spirals, the Milky Way is
warped and flared in its outskirts. The evidence comes from gas
tracers such as HI \citep{Henderson1982, Burton1986, Burton1988,
Diplas1991, Nakanishi2003, Levine2006} or molecular clouds
\citep{Grabelsky1987, Wouterloot1990, May1997}. Recent analysis of
the outer Galaxy either from 2MASS
\citep{LC2002,Momany2006,Reyle2009} or from SDSS
\citep{Hammersley2011} has led to the conclusion that even the
stars follow a warped structure. However, it has not been clearly
established that the shape of the warp is similar or deviates from
the gas structure. It is now established that the Galaxy flare is
evenly populated by young stars, as recently discovered by
\cite{Feast2014} from the Cepheids (less than 130 million years
old) in the outer disc flare at 1-2 kpc from the plane.
\cite{Kalberla2014} have compared the HI distribution with stellar
distribution (2MASS, SDSS, SDSS-SEGUE, pulsars, Cepheids) from
several authors. They argue for a typical flaring of gas and stars
in the Milky Way. \cite{Abedi2014} explored the possibility of
determining the warp shape from kinematics in the future Gaia
catalogue.

The existence of the warp can originate from perturbations of the
Galaxy by the Magellanic Clouds \citep{Weinberg2006}.
\cite{Perryman2014} argue that the tilt of the disc may vary with
time, invoking four reasons for this: i) the combination of the
infall of misaligned gas \citep{Shen2006}; ii) the interaction of
the infalling gas with the halo \citep{Roskar2010}; iii) the
effect of the Large Magellanic Cloud \citep{Weinberg2006}; iv) the
misalignment of the disc with the halo \citep{Perryman2014}.

In this work, we have investigated the shape of the outer Galactic
disc, considering its structural parameters, such as scale length,
but also the non-axisymmetric part, as warp, flare, and disc
truncation, as well their dependence with age. The inner Galaxy
and spiral structure will be presented in a forthcoming paper. To
determine accurate estimates of Galactic disc structure
parameters, we compare colour-magnitude diagrams observed with
2MASS with the predicted ones by a population synthesis model for
the Galactic plane towards the second and third quadrants at
$|\emph{b}| \le 5.5^\circ$. In the Besan\c{c}on Galaxy Model the
stellar ages serve as the driving parameters for stellar
evolution, metallicities, scale heights and kinematics. Hence, it
is  the most useful model to investigate the time dependence of
the Galactic structure parameters. The parameter fitting is done
using a powerful method for global optimisation called genetic
algorithms (GA). The GAs have been extensively employed in
different scientific fields for a variety of purposes. The main
strategy consists of adjusting the parameters of the thin disc
population, such as scale length, warp, flare, disc edge, in order
to reproduce the star counts observed by 2MASS.

The thin disc region can suffer from crowding in certain regions,
from interstellar extinction and from clumpiness which could be
difficult to model with simple assumptions. However, it is
possible to model the extinction in 3D appropriately
\citep{AL2005,Marshall2006}, such that this effect is taken out in
the analysis of the Galactic plane stellar content.

The remainder of this paper is organised as follows. In Section 2
we present the sub-sample of 2MASS data used in the present work.
In Section 3 we describe the properties of the population
synthesis model (its parameters adjusted in the current work) and
we compare the standard model with 2MASS data. The basic concepts
of the GA method and its implementation in our study are described
in Section 4. Analysis of scale length and its dependence on age,
the warp and flare shapes are presented in Section 5. The overall
discussion of Galactic structure parameters and its parameters are
presented in Section 6. In Section 7 we address the conclusions of
this study and give some final remarks.

\section{The data}

The 2MASS survey \citep{Skrutskie2006} represents the most
complete (regarding spatial coverage) and homogeneous data set in
the near infrared (in the $J$, $H$ and $K_{\rm s}$ bands)
available for the entire Galactic plane. We have performed a
selection on the 2MASS data based on the 2MASS Explanatory
Supplement using selections proposed and discussed by
\cite{Cambresy2003}. The sources to be accepted should satisfy the
following criteria: i-) the photometry uncertainty $\sigma$ $\le$
0.25; ii-) signal-to-noise ratio is larger than seven in at least
one band; iii-) contamination of extended sources must be avoided;
iv-) the field of photometry quality (\emph{Qflag}) must be
different from: \emph{X} (no possible brightness estimate),
\emph{U} (source not detected in the band or not resolved),
\emph{F} (poor estimate of the photometric error), \emph{E}
(quality of photometric profile is poor); v-) the read flag must
be different from zero, four, six, or nine. Those values indicate
either non-detections or poor quality photometry and positions.

In order to adjust parameters towards the outer Galaxy, we have
used fields located at 80$^\circ \le \ell \le 280^\circ$ and $|b|
\le 5.5^\circ$ distributed every 3\deg~and 10\deg~in longitude for
$|b|~\le 3.5$\deg~and $|b| >$ 3.5\deg, respectively. In total,
there are 2228 fields. We simulated square fields with an area
equal to $0.25 \times 0.25$ square degree in order to account for
the changes in extinction at this scale. Then to obtain higher
statistics, we grouped four fields in latitude at any given
longitude. The final fields have a size of 0.25\deg~in longitude
and 1\deg~in latitude, totalizing 557 fields.

In each field, the completeness limit of the observed sample is
estimated, and fainter stars are discarded. To obtain the
completeness limits, distributions of star counts as a function of
magnitude are built for each filter with bin size equal to 0.2
mag. The bin before the peak gives the respective completeness
limit. The source also needs to satisfy either the nominal
completeness limits of 2MASS in $J$ (15.8 mag) and $K_{\rm s}$
(14.3 mag) bands or the completeness for a given field and also to
be detected in $J$ and $K_{\rm s}$ bands. We notice that following
the criteria above, the large majority of the stars in our sample
that have simultaneous detection in the $J$ and $K_{\rm s}$ bands,
are also detected in the $H$ band. Approximately 55\% and 20\% of
the fields have a completeness limit at $K_{\rm s}$ equal to 14.1
and 14.3 mag, respectively. The total number of 2MASS stars used
in the present work is 886916.

\begin{figure}[h]
\centering
\includegraphics[scale=0.42,angle=90]{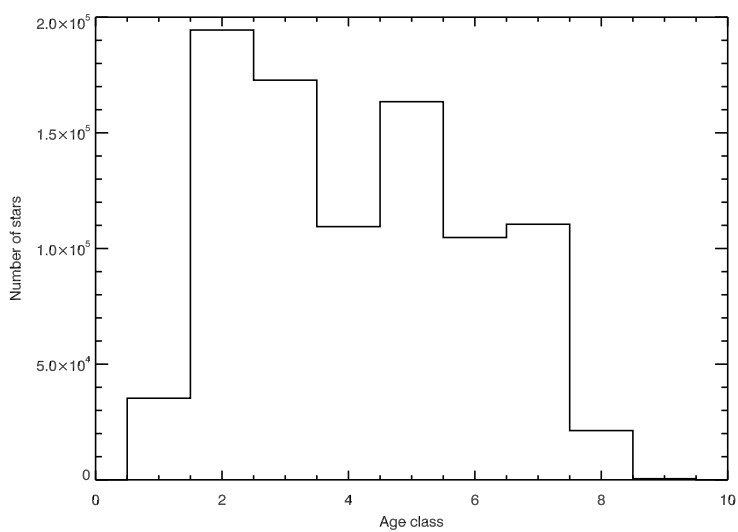}
\caption{Age distribution of simulated stars. The age classes are
provided in Table 1.} \label{nFigure1}
\end{figure}

\section{The Besan\c{c}on Galaxy Model}

To produce star counts and colour-magnitude diagrams, we make use
of the population synthesis model called Besan\c{c}on Galaxy
Model, hereafter BGM \citep{Robin1986, Robin2003, Robin2012}. It
provides a realistic description of the stellar content of the
Galaxy, assumptions about the star-formation scenario and
evolution in different populations, and includes kinematics and
dynamics as further constraints on the mass distribution. One of
the main differences between this and other Galaxy models resides
in the fact that the BGM is dynamically self-consistent locally
\citep{Bienayme1987a}. Here, we recall the main parameters
relevant for the present study. The BGM is composed of four
components: a thin disc, thick disc, halo and bar.

\begin{table}
\caption{Local mass density ($\rho_0$) and disc axis ratio
($\varepsilon$) for each thin disc sub-component of the BGM as a
function of age. AC refers to the Age Class, <\emph{age$_{i}$}> is
the mean age for each AC.}
\begin{tabular}{ccccccc}
\hline Sub-component  & Age  & <\emph{age$_{i}$}> & $\rho_0 $      & $\varepsilon$    \\
          AC          & (Gyr)& Gyr & ($M_{\odot}~ \mathrm{pc}^{-3}$)   &         \\
\hline \noalign{\smallskip}
              1& 0-0.15& 0.075 & 3.90 x 10$^{-3}$  &  0.0140 \\
              2& 0.15-1& 0.575 & 9.50 x 10$^{-3}$  &  0.0220 \\
              3& 1-2   & 1.500 & 7.53 x 10$^{-3}$  &  0.0312 \\
              4& 2-3   & 2.500 & 5.60 x 10$^{-3}$  &  0.0468 \\
              5& 3-5   & 4.000 & 7.88 x 10$^{-3}$  &  0.0598 \\
              6& 5-7   & 6.000 & 6.75 x 10$^{-3}$  &  0.0678 \\
              7& 7-10  & 8.500 & 8.20 x 10$^{-3}$  &  0.0683 \\
 \noalign{\smallskip} \hline
\end{tabular}
\end{table}

Since its first version, the BGM has been extensively compared
with several large surveys in different wavelengths and at
different depths. With regards to the on-line version
\citep{Robin2003} here we have used an update which benefits from
new results concerning the 3D extinction model
\citep{Marshall2006}, the shape of the warp and flare
\citep{Reyle2009}, the bar-bulge region \citep{Robin2012}.

Recently, a more flexible version of the model \citep{Czekaj2014}
has been proposed, which allows modification of the initial mass
function (IMF) and star-formation history (SFH) of the Galactic
thin disc. As it is still undergoing testing, its use will be
deferred to future studies. New revisions concerning the thick
disc and halo populations have been discussed by \cite{Robin2014}.
We do not make use here of the new characteristics of the thick
disc and halo, but they should affect the present study only very
marginally.

The disc population is assumed to have an age, ranging from 0 to
10 Gyr. The initial values for the evolutive parameters of the
disc (star-formation rate history, initial mass function) were
obtained by \cite{Haywood1997} from the comparisons with
observational data. The density laws for each component can be
found in Robin et al. (2003) in which the thin disc follows
Einasto laws rather than a double exponential. The disc axis ratio
of each age population, presented in Table~1, have been computed
assuming an age-velocity dispersion relation from \cite{Gomez},
using the Boltzmann equation, as explained in \cite{Bienayme1987a}
and revised in \cite{Robin2003}.

In the present work, we have adjusted parameters of the Einasto
law (scale length) and the three most important structures towards
outer Galaxy, for example, warp, flare and disc truncation. We
postpone the analysis of the inner Galaxy and spiral arms to a
further study. The scale length, warp and flare are adjusted
considering their dependence on age. While in the standard BGM the
disc is truncated at \emph{R$_\mathrm{gal}$} = 14 kpc, here we use
simulations without any truncation to be able to determine it
during the fit. The standard values for those parameters are
presented in Table 2.

\begin{table}[h]
\caption{Values of the parameters for the standard version of BGM
used in the present work. \emph{$\gamma_\mathrm{warp_{+}}$} and
$\gamma_\mathrm{warp_{-}}$ refer to the warp slope at second and
third quadrants, respectively. For scale length, kp$_{1}$ refers
to stars with age class AC $=$1, and kp$_{2...7}$ to older ages,
e.g., a unique value for the scale length of stars with AC $\geq$
2.}

\begin{tabular}{cccc}
\hline
Component & Parameter   & Unit &Value (standard BGM) \\
\hline
  warp &  $\gamma_\mathrm{warp_{+}}$ & pc kpc$^{-1}$& 0.09 \\
  &  $\gamma_\mathrm{warp_{-}}$ & pc kpc$^{-1}$ & 0.09 \\
   &  \emph{R}$_\mathrm{warp}$ &  pc & 8400 \\
   &  $\theta_\mathrm{warp}$ & rad &     0.0 \\
  flare&  $\gamma_\mathrm{flare}$& kpc$^{-1}$ &   0.05 \\
  &  \emph{R}$_\mathrm{flare}$ &  pc  &   8400 \\
scale length &  kp$_{1}$ & pc &  5000 \\
 &  kp$_{2...7}$ & pc &  2170 \\
$\chi^{2}$ & ---    &  ---  & 33.32 \\
 \noalign{\smallskip} \hline
\end{tabular}
\end{table}

\subsection{The warp and flare}

We adopt the same representation used by \cite{Reyle2009} in which
the height $z_\mathrm{warp}$ of the warp (for \emph{R} $>$
R$_\mathrm{warp}$) is defined as the distance between the
mid-plane of the disc and the plane defined by \emph{b} =
0$^\circ$. It varies as a function of Galactocentric radius
(\emph{R}) as follows.

\bigskip
\begin{equation} \label{equ:warp}
\displaystyle z_\mathrm{warp}(R) = \gamma_\mathrm{warp} \times
(R-R_\mathrm{warp}) \times \sin(\theta_\mathrm{u}-\theta_\mathrm{warp}),
\end{equation}

\bigskip

\noindent where $\theta_\mathrm{u} =
\mbox{atan2($\emph{y}$,$\emph{x}$)},$

\bigskip

\noindent and $\theta_\mathrm{warp}$ is the node angle;
$\gamma_\mathrm{warp}$ and \emph{R$_\mathrm{warp}$} are
the slopes of the amplitude and the Galactocentric radius at which
the Galactic warp starts, respectively; $\emph{x}$ and $\emph{y}$
are Galactocentric coordinates, $\emph{x}$ positive in the
Sun-Galactic centre direction, and $\emph{y}$ positive towards
rotation ($\ell = 90^{o}$).

As shown from the 2MASS star counts in \cite{Reyle2009}, the
Galactic warp acts differently on different sides (second and
third quadrants) of the Galactic disc. We have considered two
values for the warp slope for the second
($\gamma_\mathrm{warp_{+}}$) quadrant, for example, points towards
longitudes less than $\theta_\mathrm{warp}$ and third
($\gamma_\mathrm{warp_{-}}$) quadrant, for example, points towards
longitudes greater than $\theta_\mathrm{warp}$.

Concerning the flare, we  use the same representation provided by
\cite{Gyuk1999} who modelled the flare by increasing the scale
heights by a factor $k_\mathrm{flare}$ with an amplitude
$\gamma_\mathrm{flare}$, beyond a Galactocentric radius of
\emph{R$_\mathrm{flare}$}:

\bigskip
\begin{equation} \label{equ:flare}
k_\mathrm{flare}(R) = \left\{
\begin{array}{l l}

 1+\gamma_\mathrm{flare}(R-R_\mathrm{flare}),  & (\mathrm{if}  ~~  R > R_\mathrm{flare})\\
 1   & (\mathrm{if}  ~~  R <= R_\mathrm{flare}).\\
 \end{array}
  \right.
\end{equation}

\subsection{Disc truncation}

Disc truncations were first discovered in external galaxies
\citep{vanderkruit1979}. The first studies in our Galaxy were done
by \cite{Habing1988} from OH/IR stars and by \cite{Robin1992a}
from UBV photometry. Whether or not the disc is truncated, and how
the truncation scale occurs can play a role in the determination
of the thin disc scale length. The truncation can be related to
the star-formation threshold in outer galaxies
\citep{Kennicutt1989}. Due to the presence of the warp and flare,
it might happen not to be circular. \cite{Seiden1984} found
evidencee for disc truncation from the point of view of a process
of stochastic star formation.

For the gaseous component, \cite{Wouterloot1990} found a decrease
in the CO density distribution between 18 and 20 kpc (kinematic
distance method).

To model the disc truncation, we have considered a radial cut as
proposed by Robin et al. (1992b). The thin disc truncation is
computed by multiplying the density law by a factor \emph{f} which
is defined by a Gaussian truncation with a scale
\emph{h$_\mathrm{cut}$}:

\begin{equation}
f = \left\{
\begin{array}{l l}
exp(-(\frac{R-R_\mathrm{dis}}{h_\mathrm{cut}})^2) , &  \mathrm{if} R \geq R_\mathrm{dis}\\
 1. & \mathrm{if} R < R_\mathrm{dis}\\
 \end{array}
 \right.
 \label{equ:denstar_trunc}
\end{equation}

\bigskip

\noindent where \emph{h$_\mathrm{cut}$} is the truncation scale
and \emph{R$_\mathrm{dis}$} is the Galactocentric radius of disc
truncation.

\subsection{Other simulation parameters}

The simulations made by the BGM have to assume realistic
photometric errors, and the comparison should be made in a
magnitude range where the data are complete. The completeness
limit for each field was applied for two filters, $J$ and $K_{\rm
s}$, using the completeness limits obtained from 2MASS data. The
observational errors are assumed to follow an exponential law as a
function of magnitude \citep{Bertin1996} as in Eq~\ref{eq-error}.

\begin{equation}
\label{eq-error} \sigma _M =A+\exp (C\times m - B),
\end{equation}

\noindent where \emph{m} is the magnitude in the observed band,
and \emph{A}, \emph{B}, \emph{C} are parameters obtained by
fitting on 2MASS observational errors. These parameters are
computed for each 2MASS field, because of varying observational
conditions. The fit was performed on 2MASS data before cutting
them at the completeness limit.

Since the fields used in the present work are in the Galactic
plane, the interstellar extinction is crucial in the analysis of
colour-magnitude diagrams \citep{AL2005, Marshall2006}. Hence, it
is mandatory to use a model with good spatial resolution and to
estimate the reddening accurately. However the parameters which
are being studied (warp, flare, scale length, disc truncation)
affect only the density, meaning that the amplitudes and not the
shape of the histograms ($J - K_{\rm s}$), and cannot mimic a
reddening. To have the best estimate of the distribution of
extinction along every line of sight, we adopt the 3D extinction
model proposed by \cite{Marshall2006} with a posterior revision
Marshall (2009, private communication) covering the entire
Galactic plane including Galactic anti-centre regions.

\subsection{Catalogue of simulated stars}

We start the process by creating a catalogue of simulated stars
from the standard axisymmetric model. A typical simulation using
the model provides a catalogue of stars with their properties,
such as distance, colours, magnitude, age, luminosity class,
effective temperature, gravity and metallicities. We apply the
same selection function on the model simulation as on the data. In
total, there are 912417 pseudo-stars produced by the model and
886916 observed by 2MASS with the same selection.

Figure~\ref{nFigure1} shows the distribution of simulated stars
according to their age bin, see Table 1, where the age class (AC)
equal to eight corresponds to thick disc stars.

Next, we distributed the stars in colour bins of size 0.5 in
$J-K_{\rm  s}$ to obtain star counts N($J-K_{\rm s}$, $\ell$,
\emph{b}). Finally, we merged the bins in colour ($J-K_{\rm s}$)
in which the number of stars is less than 100 (either for 2MASS or
BGM) with the right neighbour bin in $J-K_{\rm s}$.

We discarded a few bins (43) in $J-K_{\rm s}$ which show large
discrepancies with their neighbours in the ($\ell$,\emph{b})
space, where we suppose that either the data suffer from errors or
the extinction is not well modelled. The final total number
(\emph{N$_{\rm bin}$} in Eqn 5) of bins is 1615. In the catalogue
of simulated stars, there are 23193 and 7292 stars for the range
of 15 kpc $\le$ \emph{R} $<$ 18 kpc and \emph{R} $\ge$ 18 kpc,
respectively, which allows us to constrain the distribution of
stars at large distances in the Galactic plane.

\subsection{Comparison between the standard model and 2MASS}

Next, to analyse the differences between the standard model with
2MASS data we computed the relative residuals in each colour bin,
as defined below.

\bigskip
\begin{equation} \label{equ:xi}
\displaystyle \epsilon_{\textrm{rel}} = \sum_{i=1}^{N_{\rm
bin}}(N_{i,\rm obs}-N_{i,\rm model})/N_{i,\rm obs},
\end{equation}
\noindent
\bigskip


\noindent in which \emph{N$_{i,\rm model}$}, and \emph{N$_{i,\rm
obs}$} are the model (BGM standard) and observed counts in the
space ($J-K_{\rm s}$, $\ell$, \emph{b}), and \emph{N$_{\rm bin}$}
is the number of bins.

\begin{figure}[ht]
\includegraphics[scale=0.35,angle=90]{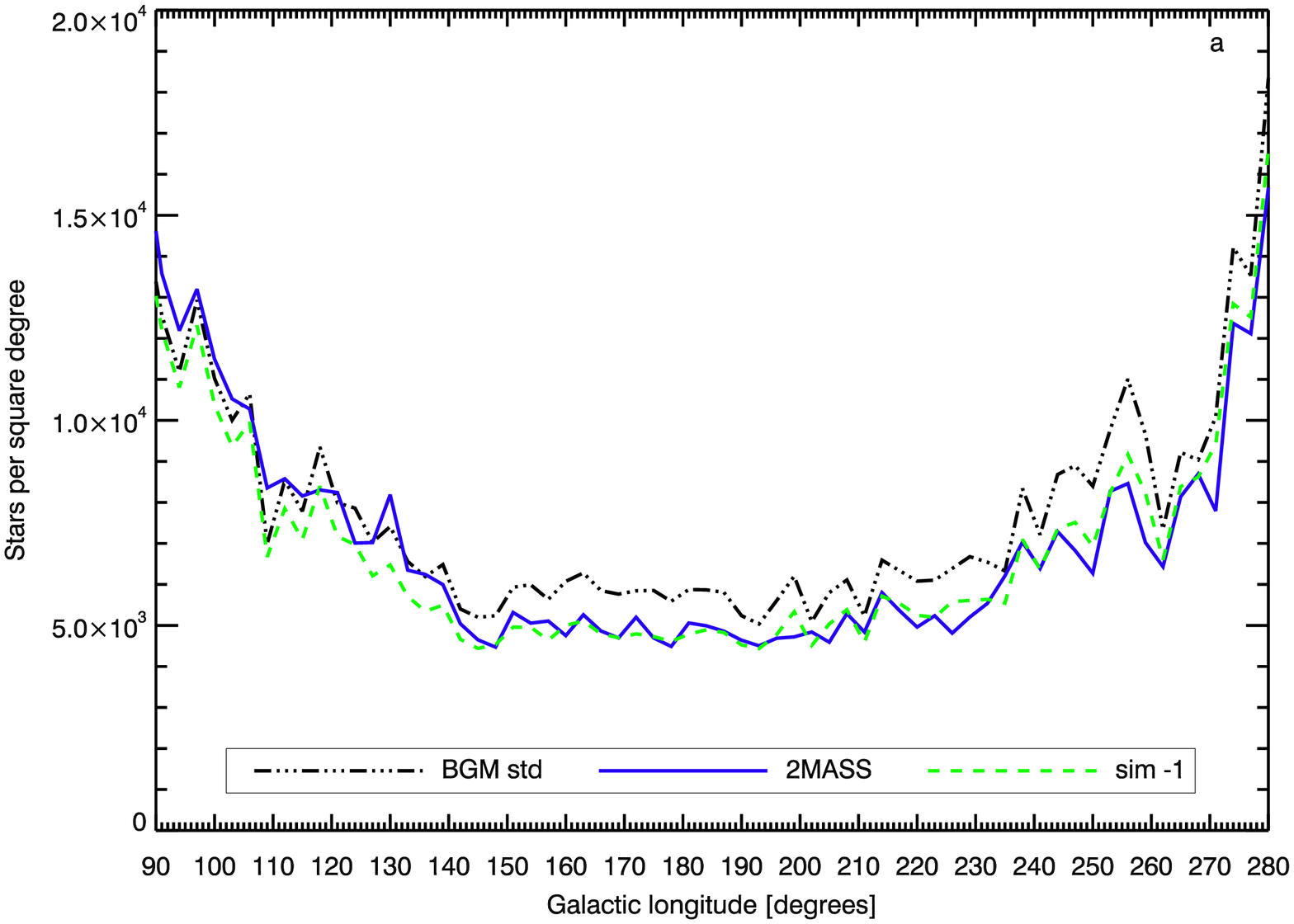}
\includegraphics[scale=0.35,angle=90]{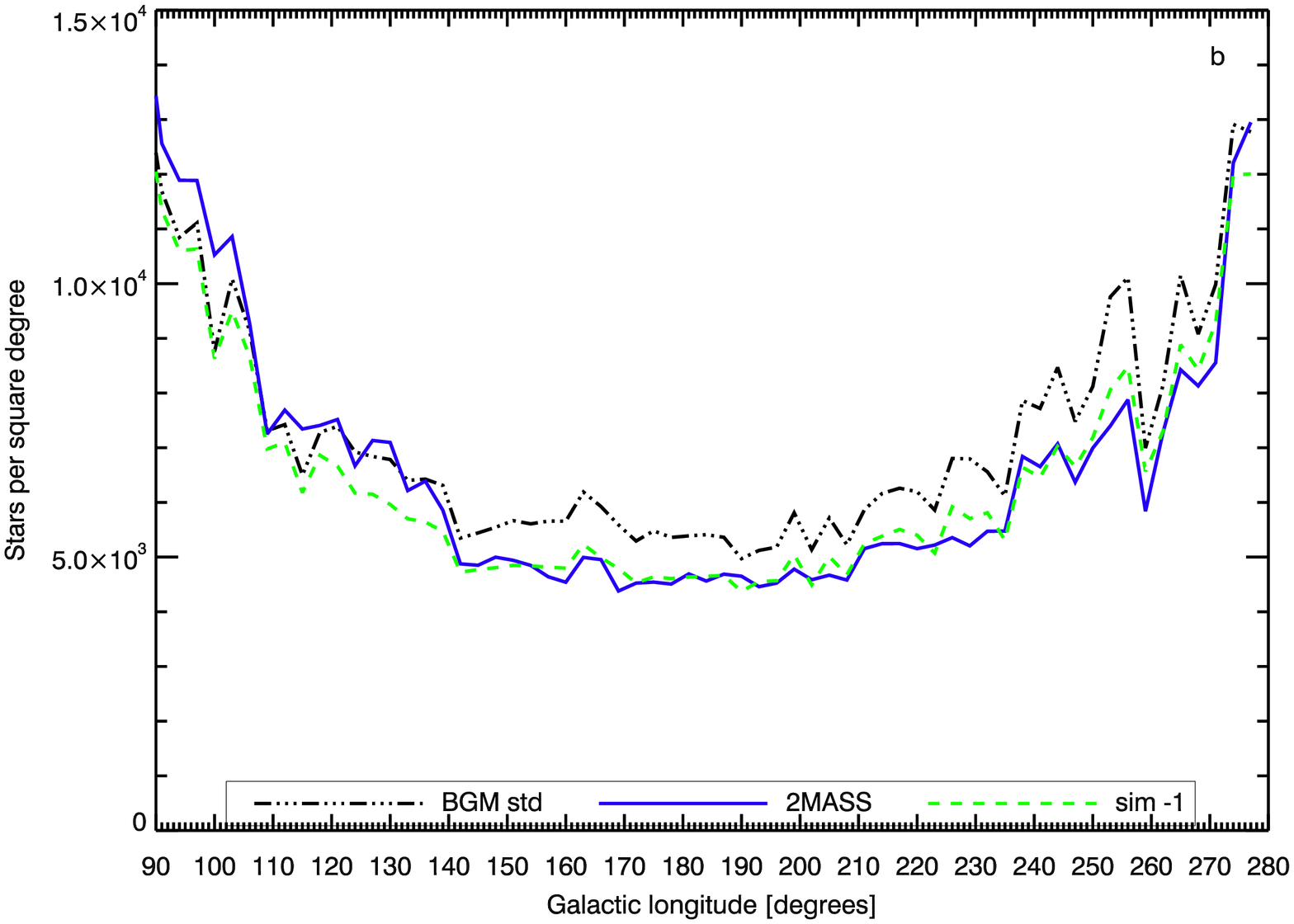}
\includegraphics[scale=0.35,angle=90]{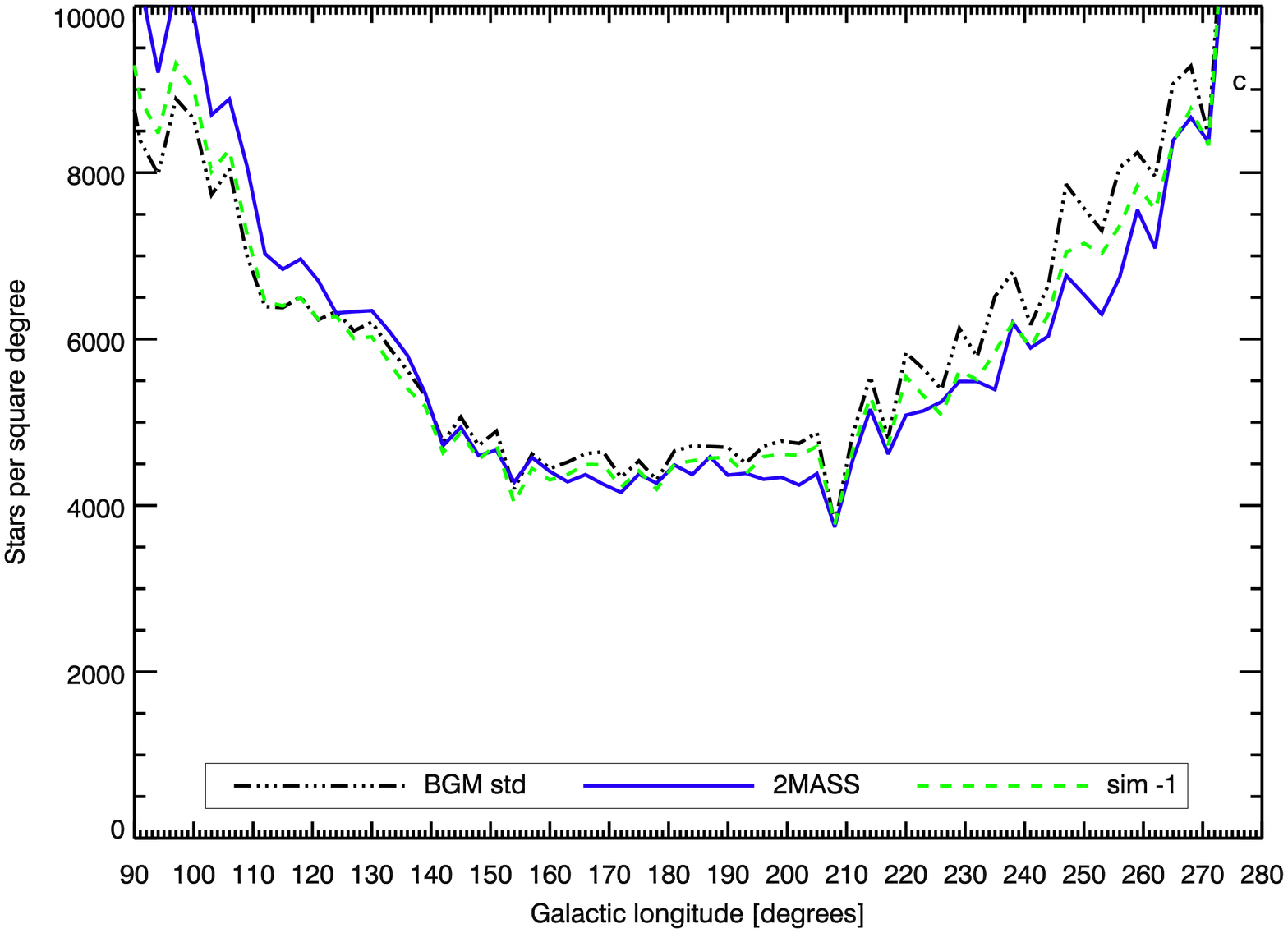}
\centering \caption{Longitudinal profile for star counts for three
ranges of Galactic latitudes. a) $|b| \leq 0.5^\circ $; b) $0.5
^\circ < |b| \le 3.5^\circ$; c) $|b|
> 3.5^\circ$. The data are plotted in solid lines, the standard
BGM with dotted-dashed lines, sim-1 in dashed lines (see text).
The Orion spiral tangents are located at approximately at $\ell
\sim 80$ and 260$^{\circ}$.} \label{lgtdprofiles}
\end{figure}

\begin{figure*}[ht]
\centering
\resizebox{14.0cm}{!}{\includegraphics[angle=90,scale=.90]{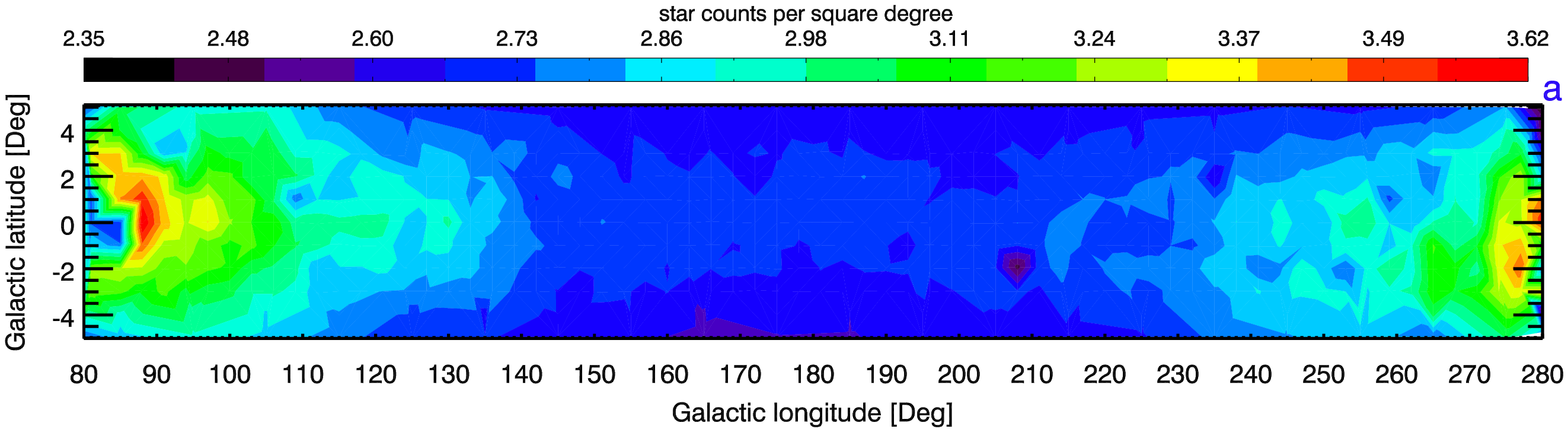}}
\resizebox{14.0cm}{!}{\includegraphics[angle=90,scale=.90]{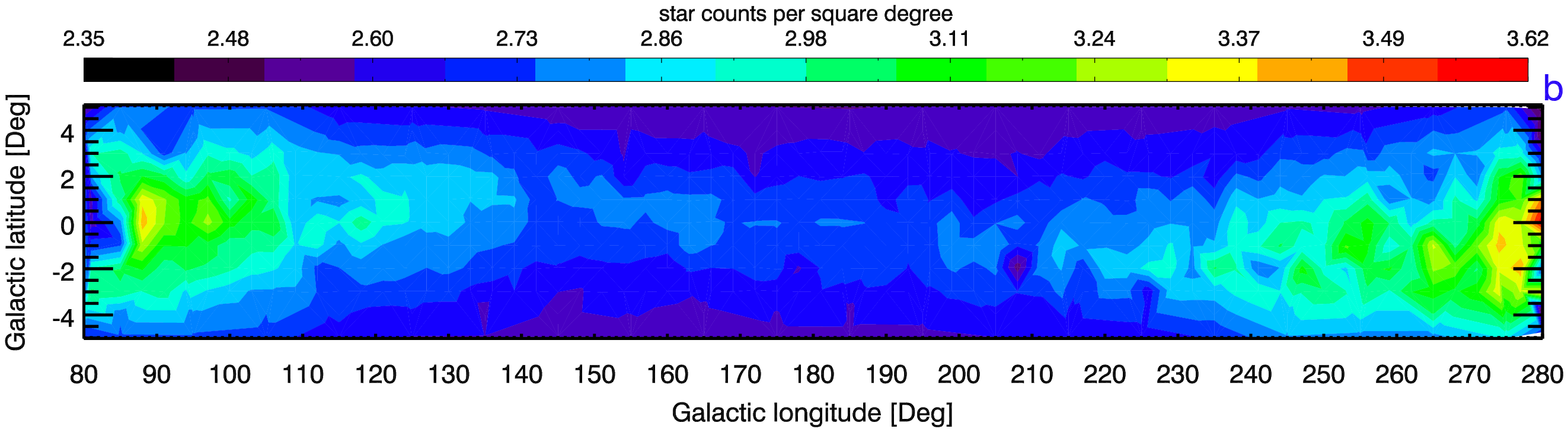}}
\resizebox{14.0cm}{!}{\includegraphics[angle=90,scale=.90]{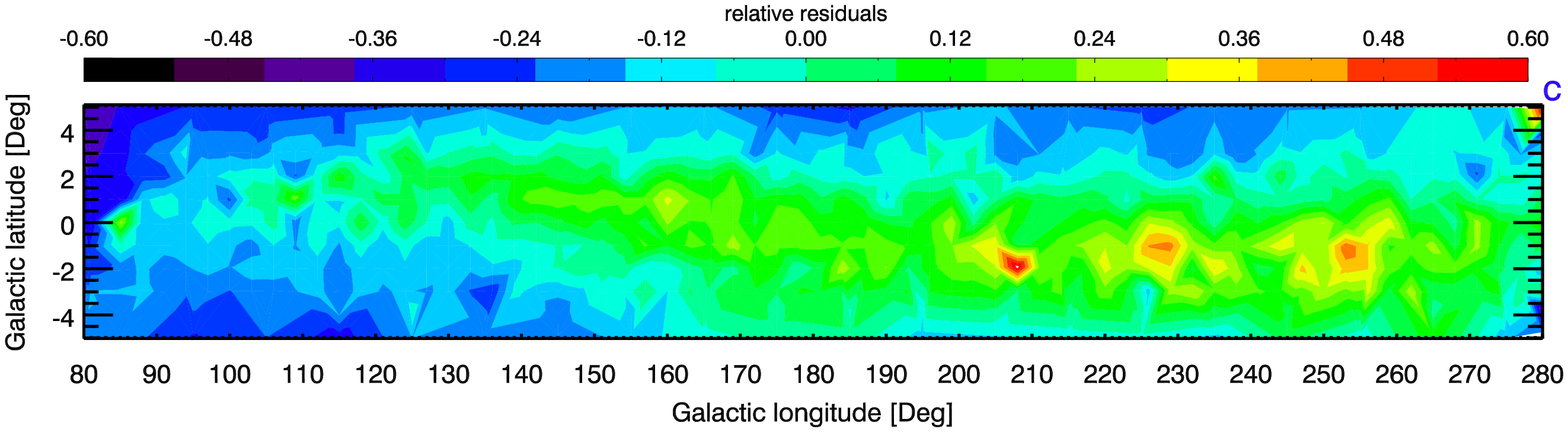}}
\resizebox{14.0cm}{!}{\includegraphics[angle=90,scale=.90]{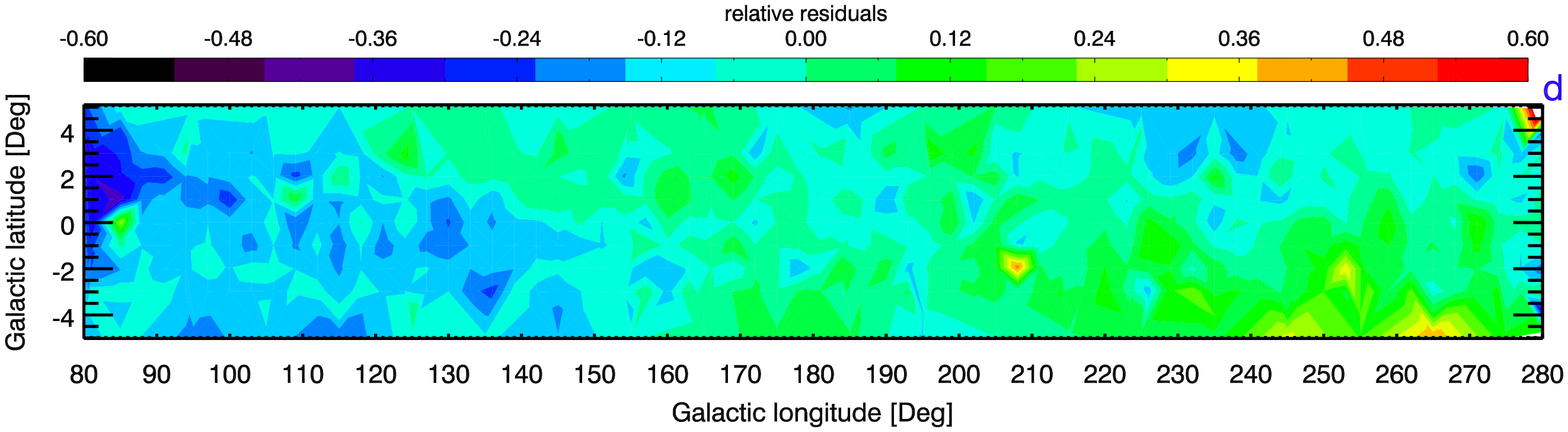}}
\caption{Maps of star counts per square degree and residuals of
the fit for standard model and sim-1: a) observed 2MASS star
counts; b) standard model star counts; c) relative residuals for
standard model; d) relative residuals for sim-1 (Eqn 5). The
values are binned ($\Delta \ell = 3^\circ$ and $\Delta
b=0.5^\circ$) and interpolated. The Orion spiral tangents are
located at approximately at $\ell \sim$ 80 and 260$^{\circ}$.}
\label{maps_std_sim1}
\end{figure*}

The differences in star counts as a function of longitude between
the standard model and 2MASS data are presented in
Fig.~\ref{lgtdprofiles} and as a map of relative residuals in
Fig.~\ref{maps_std_sim1}.

\section{The optimisation procedure}

In order to fit model parameters, we have used a genetic
algorithm, a method based on evolutionary mechanisms and theories,
using selections of models similar to natural selection and
genetics. This method presents some characteristics which make
this technique more efficient than the usual heuristic methods
based on calculus, either random or enumerative processes as
pointed out by \cite{Holand1975} and \cite{Mitchell1996}, and
references therein. The GAs are also related to artificial
intelligence, having the ability to learn by experimenting, as
used in many computational domains.

In astronomy, different versions of the GA have been employed in
several different applications, involving Galaxy modelling, such
as \cite{Sevenster1999} and stellar population diagnostics by
colour-magnitude diagrams \citep{Ng1998, Ng2002}. In another
instance, \cite{Larsen2003} applied the GA to retrieve eight
parameters of the Galactic structure. They performed comparisons
between their model counts and the data from the Automated Plate
Scanner Catalog for 88 fields.

In the present work, we have used the version of GA called
PIKAIA\footnote{available at \url{
http://www.hao.ucar.edu/modeling/pikaia/pikaia.php}}. This
optimisation subroutine was first presented by \cite{Charb1995}
and \cite{Charbonneau1996}. PIKAIA works with 12 parameters
\citep{Charb1995}. The choice for the values of those parameters
depends on the application. After some tests, we chose the values
of these parameters described in Table 3. Some of them are
slightly different from the standard ones; they are appropriate
when a large number of parameters are adjusted. They force a
higher search in the space of parameters, as for instance, the
crossover probability and the maximum mutation rate.

The \textit{ngen} parameter defines the number of the generations
used. In our case, a too-small value would cause a premature
solution. On the other hand, we identified that for our problem,
there is no significant improvement in the $\chi^{2}$ with
\emph{ngen} values of greater than 300.

\bigskip

\begin{table*}[t]
\centering \caption{Values of the parameters set in PIKAIA with
their meaning in comparison with the standard ones.}
\begin{tabular}{llll}
\hline  Parameter &Present work & Default & Identifier name \\
\hline
\emph{ng}  & 100   & 100   & number of individuals in a population  \\
\emph{ngen} & 300   & 500   &  number of generations over which solution is to evolve \\
\emph{nd}  &  5   & 6     &  number of significant digits  \\
\emph{crossover}  & 0.90  & 0.85  & crossover probability \\
\emph{mut}& 2   &  2    &  mutation mode: 1 - one-point mutation,
fixed rate; 2 - one-point, adjustable rate based on fitness; \\
 &    &       &  3 - one-point, adjustable rate based on distance; 4 - one-point + creep, fixed rate; \\
 &    &       &  5 - one-point+creep, adjustable rate based on fitness; \\
 &    &       &  6 - one-point+creep, adjustable rate based on distance \\
\emph{imut}  & 0.005 & 0.005 & initial mutation rate \\
\emph{pmutmn} & 0.0005&0.0005  &  minimum mutation rate\\
\emph{pmutmx}  & 0.35  &0.25   & maximum mutation rate\\
\emph{fdif}  & 1.0   &1.0    &  relative fitness differential\\
\emph{irep}  & 1     &3      &  reproduction plan: 1 - full generational; 2- steady state replace-random;  \\
  &    &       &  3 - state-replace-worst (only with no elitism) \\
\emph{ielite}  & 1     &0      & elitism: 1=on, 0=off\\
\noalign{\smallskip} \hline
\end{tabular}
\end{table*}

The \textit{nd} parameter defines the number of significant digits
retained in chromosomal encoding \citep{Charb1995}. For the
parameters used in the present work, a value of 5 for $nd$ was
found to be the most reasonable value as a compromise between
accuracy and performance. Indeed a run with \emph{nd} = 6 gave
similar parameter values and $\chi^{2}$. Regarding CPU time, the
difference was approximately 10\% larger than with \emph{nd} = 5.

We also have used the elitism technique that consists of storing
away the parameters of the best-fit member of the current
population and later copying them into the offspring population
\citep{Charb1995}. Use of this technique also avoids a lost
solution of good gens by either mutation or crossover. As we have
used elitism, it is necessary to use the options (\emph{irep}) 1
or 2 (see Table 3) in the reproduction plan. We have used the
option \emph{irep} = 1 (full generational replacement). We also
did a run using \emph{irep} = 2, but the $\chi^{2}$ was similar,
as well as the parameters, considering the range of standard
deviation.

The main procedure consisted of comparing the counts obtained from
drawn parameters with the observed data, and making the parameters
evolve to improve the figure of merit. In order to optimise the
process and to avoid recomputing all simulations each time the
model parameters are changed, we attribute to each star ($s$) of
coordinates (\emph{x},\emph{y},\emph{z}) a weight ($w_\mathrm{s}$)
which is the ratio between the new density
($\rho_\mathrm{new}(x,y,z)$) obtained by parameter fitting and the
standard one ($\rho_\mathrm{std}(x,y,z)$) obtained with BGM
standard parameters, as described below:

\bigskip
\begin{equation} \label{equ:denstar_thin}
\displaystyle w_\mathrm{s}  =
\frac{\rho_\mathrm{new}(x,y,z)}{\rho_\mathrm{std}(x,y,z)},
\end{equation}
\noindent
\bigskip

This was only applied to the thin disc population. Other
populations were unchanged in the process. Positions are given as
a function of Galactocentric cartesian coordinates
(\emph{x},\emph{y},\emph{z}).

\bigskip

The total number of stars (\emph{N$_{i,\rm new}$}) modelled (for
the given set of parameters) for each bin (\emph{i}) is given by
Eqn~\ref{equ:totbi}:

\bigskip
\begin{equation} \label{equ:totbi}
\displaystyle N_{i,\rm new} = \sum_{s=1}^{N_{i,\rm std}}
w_\mathrm{s}.
\end{equation}
\noindent
\bigskip

\noindent \emph{N$_{i,\rm std}$} is the number of stars in the bin
(\emph{i}) in the standard model.

\bigskip

The merit function is presented in Eqn~\ref{equ:xi}:

\bigskip
\begin{equation} \label{equ:xi}
\displaystyle \chi^{2} =
\sum_{i=1}^{N_{\rm bin}}(N_{i,\rm obs}-N_{i,\rm new})^{2}/(N_{i,\rm obs}+N_{i,\rm new})^{2},
\end{equation}
\noindent
\bigskip

\noindent \emph{$N_{i,\rm obs}$} is the number of stars in bin
(\emph{i}) observed by 2MASS, \emph{N$_{\rm bin}$} is the number
of bins (1615) and \emph{N$_{i,\rm new}$} defined in Eqn 7.

We preferred to use this relation instead of traditional
$\chi^{2}$ in order to have a reasonable weight for bins with a
large number of stars. We note that a relative $\chi^{2}$ avoids
overweighting the contribution of latitude bins with high star
counts but without necessarily high contrast. \cite{Larsen2003}
used a similar one. As a reference, the total $\chi^{2}$ of the
standard model is 33.32 for 1615 bins, distributed in 14.70
towards second quadrant (784 bins) and 18.62 towards third
quadrant (831 bins). Table 4 shows the range of parameters
involved in those simulations.

\begin{table}[h]
\caption{The range of the parameter values used in sim-1.}
\begin{tabular}{cccc}
\hline
Component & Parameter &  Unit & Range \\
\hline
  warp      &  $\gamma_\mathrm{warp_{+}}$ & pc kpc$^{-1}$ & [0.01;0.81] \\
            &  $\gamma_\mathrm{warp_{-}}$ & pc kpc$^{-1}$ & [0.01;0.81] \\
            &  $R_\mathrm{warp}$          & pc            & [7000;12000]  \\
            &  $\theta_\mathrm{warp}$     & rad           & [0;$ 2\pi$]  \\
  flare&  $\gamma_\mathrm{flare}$         & kpc$^{-1}$    & [0.005;0.055] \\
            &  $R_\mathrm{flare}$         & pc            & [8000;11000]  \\
scale length&  kp$_{1}$                   & pc            & [3500;6500] \\
            &  kp$_{2...7}$               & pc            & [1200;3700] \\
disc truncation  &  \emph{R$_\mathrm{dis}$}&pc            & [12000;22000] \\
            &  \emph{h$_\mathrm{cut}$}      & pc & [500;1500]  \\
 \noalign{\smallskip} \hline
\end{tabular}
\end{table}

\section{Results}

\subsection{Fitting the parameters of the standard model}

The standard BGM works with two scale lengths for disc stars,
first one, kp$_{1}$, for stars with age $<$ 0.15 Gyr and the
second one, kp$_{2...7}$, for stars ranging from 0.15 to 10.0 Gyr
(see also Table 1). We performed a set of fitting procedures
(sim-1) with 100 independent runs of 300 generations each,
considering these two scale lengths for disc stars, and warp and
flare parameters. Table 5 summarises the fitted parameters in
sim-1, as well as the median and standard deviation for each
parameter.

\begin{table}[ht]
\centering \caption{Parameters obtained for sim-1: median and
standard deviation for 100 independent runs. A unique scale length
is considered for age class 2 to 7, noted by kp$_{2...7}$. $Lr$ is
the reduced likelihood and BIC is the Bayesian information
criterion, see Eqn 16 and text in Section 5.2.}
\begin{tabular}{ccc}
\hline
   Parameter & Unit & sim-1 \\
\hline        $\gamma_\mathrm{warp_{+}}$  & pc kpc$^{-1}$ & 0.626 $\pm$ 0.047 \\
              $\gamma_\mathrm{warp_{-}}$  & pc kpc$^{-1}$ & 0.165 $\pm$ 0.021 \\
              $R_\mathrm{warp}$           & pc            &  9108 $\pm$ 144\\
              $\theta_\mathrm{warp}$      & rad           & 3.292 $\pm$ 0.024\\
              $\gamma_\mathrm{flare}$     & kpc$^{-1}$    &  0.23 $\pm$ 0.04\\
              $R_\mathrm{flare}$          & pc            &  8923 $\pm$  191\\
              kp$_{1}$                    & pc            & 3852 $\pm$  167\\
              kp$_{2...7}$                & pc            & 2477 $\pm$  48\\
              {\emph{R$_\mathrm{dis}$}}                   & pc            & 16081 $\pm$ 1308\\
              {\emph{h$_\mathrm{cut}$ }}                  & pc            &  717 $\pm$ 272 \\
           $\chi^{2}$                     & ----------    &  21.80 $\pm$ 0.08\\
            $Lr$                          & ----------    &   -20029.8       \\
            BIC                           & ----------    & 40133.4         \\
\noalign{\smallskip} \hline
\end{tabular}
\end{table}

The scale length for stars with age larger than 0.15 Gyr is best
fitted by a scale length of 2.48 kpc. This value is similar to the
scale length used in the standard BGM, and also in agreement with
many recent studies (see discussion in Section 6). As shown in
Table~5, the warp is found asymmetrical, with a stronger slope in
the third quadrant than in the second quadrant. The flare is also
very significant. The starting radius of the flare is close to the
starting radius of the warp.

Figure~\ref{lgtdprofiles} shows longitudinal profiles for sim-1
versus BGM standard and 2MASS data for three ranges of Galactic
latitudes. There is a net improvement with the new fits at nearly
all longitudes.

Depending on latitudes, the sim-1 model performs better or
similarly to the standard BGM. For example, the agreement is
better at $|\emph{b}|<$ $0.5^\circ$, apart from the longitude
$\ell < 100^\circ$ where it is similar. Though, at the same
longitudes but higher latitudes (\emph{b} $>$ 3.5$^\circ$) sim-1
is much better. Overall, sim-1 performs much better than the
standard model, although it is not perfect everywhere. We expect
that a more complex model might give a better solution (see
Section 5.2).

Indeed, part of  the differences at $\ell \sim 80^{\circ}$ could
be attributed to the Local Arm \citep{Drimmel2000} that is not
modelled in the present work. He also pointed out that the
counterpart of the Local arm is centred at $\ell \sim
260^{\circ}$, where we also see some disagreement between our
model and 2MASS data. The fitting of spiral arms using BGM with
2MASS data is outside of the scope of the present paper. It will
be analysed in detail in a forthcoming paper where we attend to
fit a spiral model including data from the inner Galaxy.

The irregular profile can be due to the interval of bins in
longitude and because each field has its extinction, photometric
errors and completeness limits. Another source of error resides on
the fact that the map of Marshall (2009) has a lower resolution
for outer Galaxy fields than for inner part. For instance in some
fields, we got extinction determination for a distance of
approximately 0.5 degrees of the centre of our field.

Figures~\ref{maps_std_sim1}a and~\ref{maps_std_sim1}b show the
counts observed by 2MASS and modelled by BGM Standard,
respectively. Figures~\ref{maps_std_sim1}c and
Figures~\ref{maps_std_sim1}d show the ($\ell$,\emph{b}) map of the
relative residuals for the standard model and sim-1, respectively.
In Fig.~\ref{maps_std_sim1}c, some regions in orange, red, yellow
show where the standard model predicts more counts than observed.
An interesting pattern is that the position of the mid-plane
slightly differs in the model from the data. This is particularly
the case in the second quadrant between $\ell \sim$ 120$^\circ$
and 180$^\circ$. In the third quadrant, the mid-plane seems to be
almost correctly placed but, the model overestimates the counts at
latitudes $\sim0 ^\circ$. A similar feature, due to dissymmetry of
the disc in the warp and probably in disc truncation, has been
pointed out by \cite{Reyle2009}. They show that the warp slope was
different in the second and third quadrants. While the differences
towards $\ell \sim 90^\circ$ and 270$^\circ$ can be attributed
mainly to the warp, the overdensities found in Galactic
anti-centre are probably due to the disc scale length, disc edge
and flare effects, as it is investigated below.

In Fig.~\ref{maps_std_sim1}d, in the comparison of sim-1 with
2MASS data, we can identify that most of the fields for both
second and third quadrants have a relative difference in the range
of 0$-20\%$. A minuscule area can be seen at ($\ell$,\emph{b})
$\sim$ (210$^\circ$, -2$^\circ$). We note that in analysing the
Marshall (2009) extinction map, this region has a high extinction
with A$_{V}$ of about 10 to 12 mag, while in the neighbouring
fields the extinction drops to four magnitudes in A$_{V}$. This
explains the reduced counts not only in the model counts but also
in 2MASS data as can be seen in Fig.~\ref{maps_std_sim1}a. This
high extinction area is not well taken into account in the 3D
extinction model used.

\begin{figure}[h] \centering
\includegraphics[scale=0.37,angle=90]{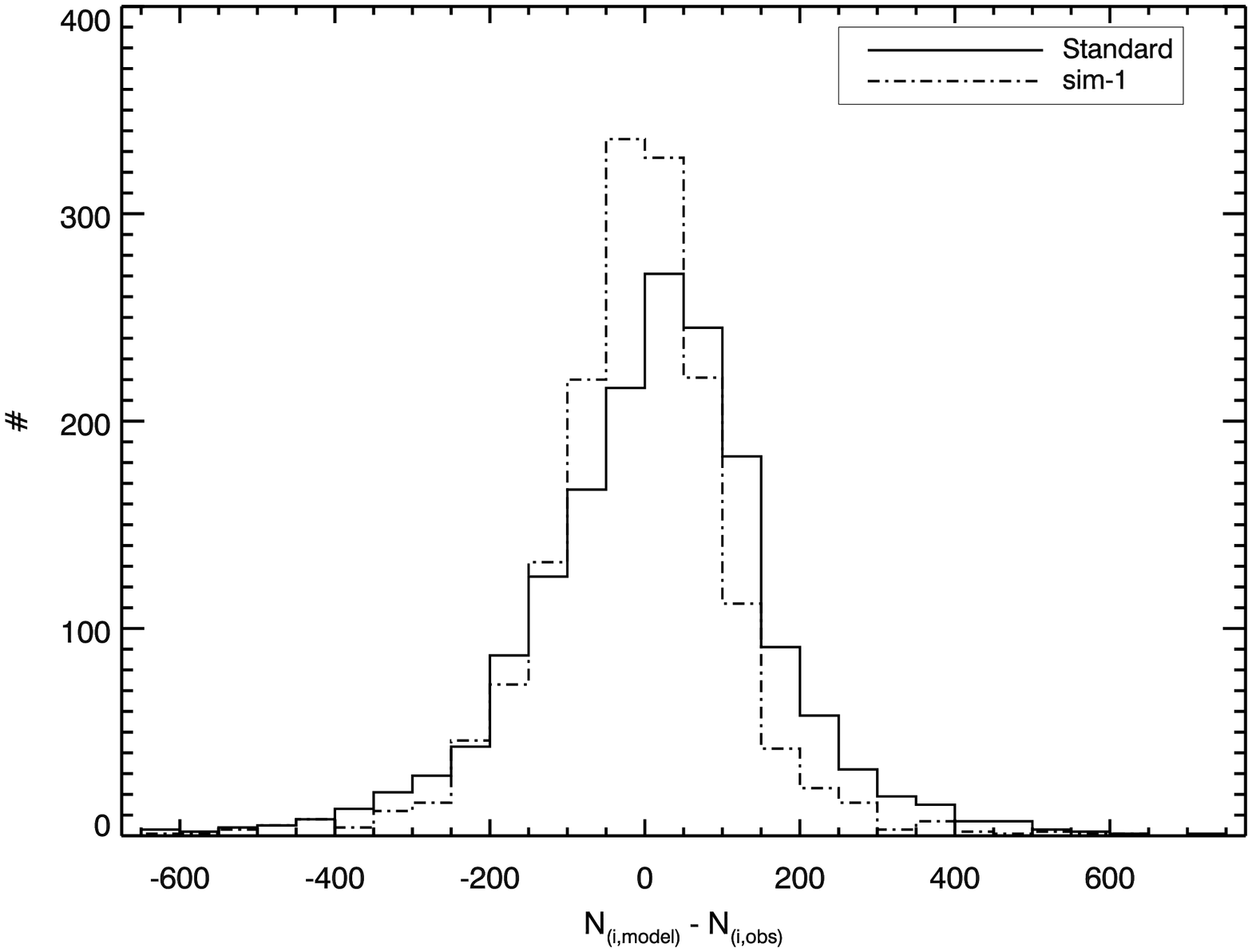}
\includegraphics[scale=0.37,angle=90]{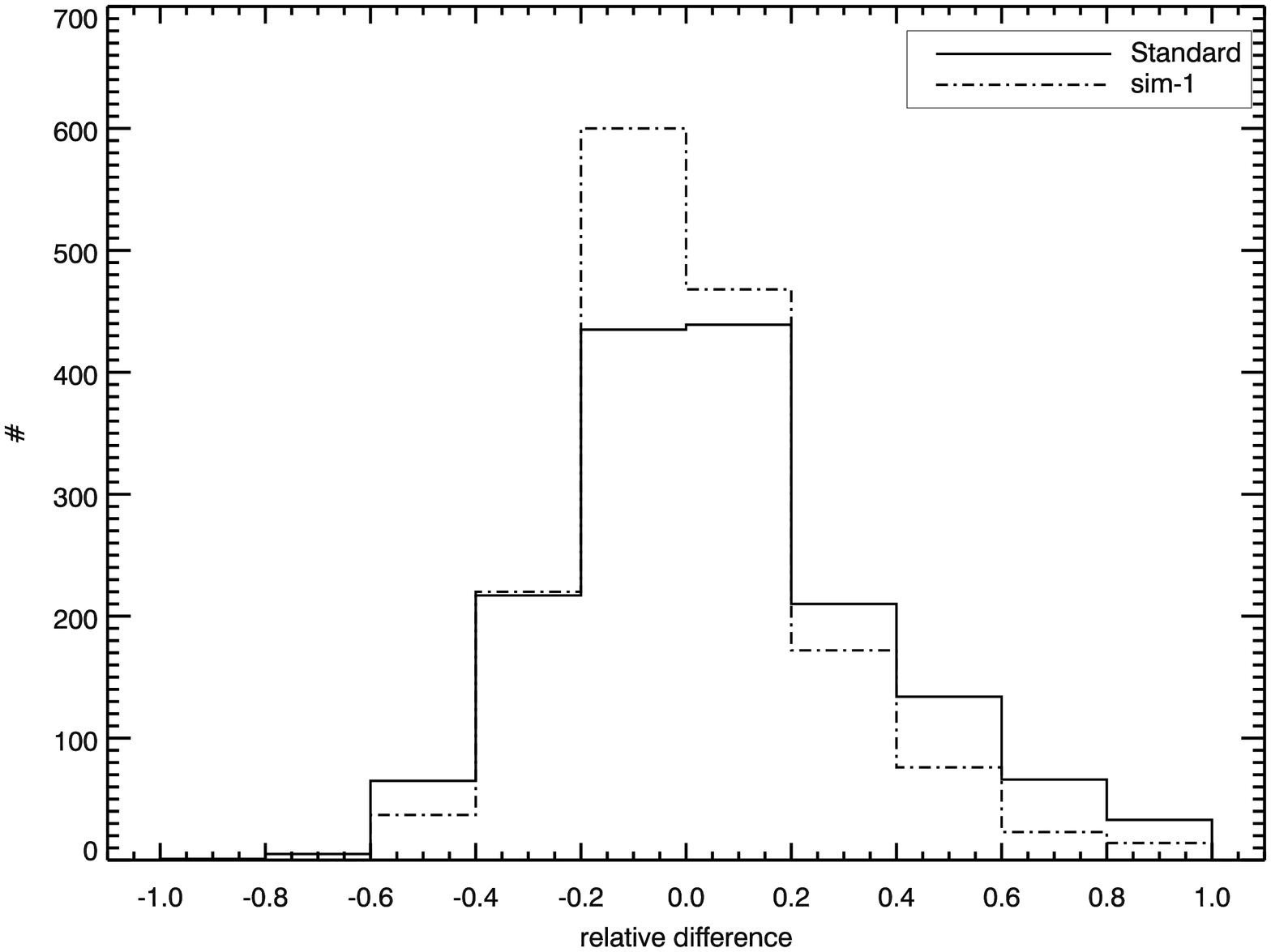}
\centering \caption{Histogram of the differences. Upper: Absolute
difference, $N_{i,\rm model}-N_{i,\rm obs}$ per bin; Lower:
relative difference. Standard model (solid line), sim-1
(dotted-dashed line).} \label{histSim1}
\end{figure}

It can also be seen that both second and third quadrants present a
significant improvement with regards to the standard model. We
note that it is still better in the third quadrant, apart from the
region of the local arm. Some residuals appear larger than 10\% at
$\ell \sim$ 140$^\circ$, which could be related to the outer arm,
not included here. After the optimisation procedure, most of the
fields with green colour, for example, relative differences around
30-40\%, decreases to residuals smaller than 20\%.
Figure~\ref{histSim1}~(upper panel) shows the histogram of the
differences $N_{i,\rm model}-N_{i,\rm obs}$ per bin for the
standard model and sim-1. As can be seen in
Figure~\ref{histSim1}~(lower panel), there is a significant
increase of bins with a relative difference of less than 20\% in
sim-1.

\begin{figure*}[ht]
\centering
\includegraphics[height=15.5cm,angle=90.0]{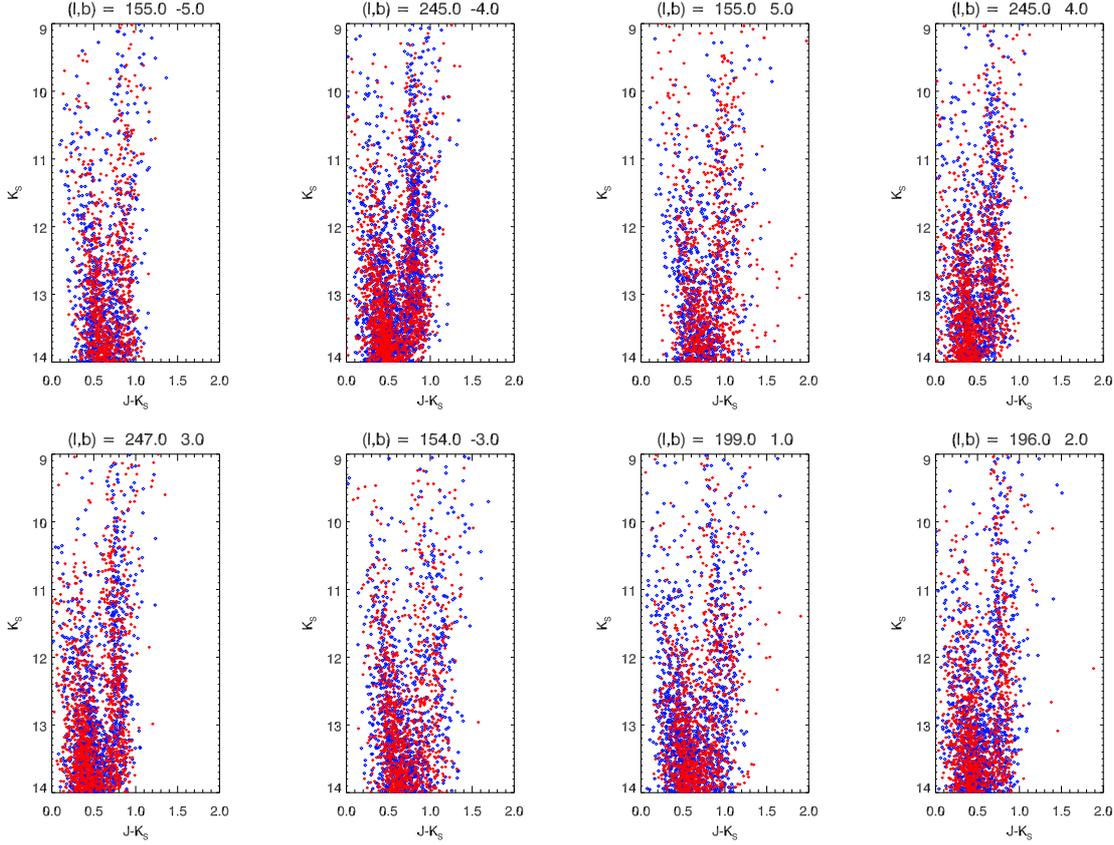}
\caption{Colour-magnitude diagrams~($K_{\rm  s}$, $J-K_{\rm s}$)
for eight directions with their coordinates (in degrees) presented
at the top of each panel: red diamonds representing 2MASS data and
blue diamonds representing BGM Standard.} \label{cmds}
\end{figure*}

In order to investigate whether the discrepancies between
simulations and data can be attributed to the interstellar
extinction, we have analysed eight colour-magnitude diagrams
($K_{\rm s}$, $J-K_{\rm s}$) towards the regions with substantial
differences in the modelled counts. Figure~\ref{cmds} shows those
diagrams for 2MASS data and simulations with the standard model.
It is clear that the extinction cannot be invoked to interpret the
difference in star counts seen between 2MASS data and the model,
the colours being in good agreement. We note that both standard
BGM simulations and the fitted models named sim-1 and sim-2 (see
next section) use the same extinction map.

To improve the model, we have explored whether dependencies of
parameters with age can better describe the observed data. This is
presented in the next section.

\begin{table*}[t]
\centering \caption{Parameters obtained for sim-2 and sim-3 with
parameters dependent on age (expressions shown in Eqn 9 to 15).
For sim-3, we have considered the ripples (see Section 5.5). The
likelihood $Lr$ and the Bayesian Information Criterion BIC are
explained in Eqn 16 and in Section 5.2}.

\begin{tabular}{cccc}
\hline
              Parameter   & Range & sim-2 &  sim-3 \\
\hline
              $a_{\gamma_\mathrm{warp_{+}}}$ (pc kpc$^{-1}$) &  [0.0;1.30]             & 1.033 $\pm$ 0.221 & 0.993 $\pm$ 0.237\\
              $b_{\gamma_\mathrm{warp_{+}}}$ (pc kpc$^{-1}$ Gyr$^{-1})$  & [0.15;0.75] & 0.357 $\pm$ 0.139 & 0.367 $\pm$ 0.133\\
              $c_{\gamma_\mathrm{warp_{+}}}$ (pc kpc$^{-1}$ Gyr$^{-2})$  & [0.02;0.12] & 0.053 $\pm$ 0.024 & 0.056 $\pm$ 0.024\\
              $a_{\gamma_\mathrm{warp_{-}}}$ (pc kpc$^{-1}$) & [0.05;0.40]      & 0.062 $\pm$ 0.031 & 0.074 $\pm$ 0.039\\
              $b_{\gamma_\mathrm{warp_{-}}}$ (pc kpc$^{-1}$ Gyr$^{-1})$  & [0.06;0.26]      & 0.082 $\pm$ 0.019 & 0.083 $\pm$ 0.019\\
              $a_{R_\mathrm{warp}}$          (kpc) &  [7000;12000]  & 10189 $\pm$ 395   & 10510 $\pm$ 446\\
              $b_{R_\mathrm{warp}}$          (kpc Gyr$^{-1})$ & [100;600]  &   255 $\pm$  74   &   301 $\pm$  80\\
              $a_{\theta_\mathrm{warp}}$ (rad) & [2.50;3.50]       & 3.382 $\pm$ 0.056 & 3.376 $\pm$ 0.058\\
              $b_{\theta_\mathrm{warp}}$ (rad Gyr$^{-1})$ & [0.03;0.23]      & 0.058 $\pm$ 0.012 & 0.055 $\pm$ 0.011\\
              $a_{R_\mathrm{flare}}$                (kpc) & [7000;11000]  & 8024  $\pm$ 316   & 7992  $\pm$ 324\\
              $b_{R_\mathrm{flare}}$                (kpc Gyr$^{-1})$ & [300;1300]  &  922  $\pm$ 193   &  929  $\pm$ 239\\
              $a_{\gamma_\mathrm{flare}}$           (kpc$^{-1})$& [0.15;0.65]      & 0.359 $\pm$ 0.121 & 0.328 $\pm$ 0.117\\
              $b_{\gamma_\mathrm{flare}}$           (kpc$^{-1}$ Gyr$^{-1}$)& [0.01;0.16]     & 0.051 $\pm$ 0.039 & 0.0601$\pm$ 0.035 \\
              \emph{R$_\mathrm{dis}$}  (pc)                & [12000;22000] & 19482 $\pm$ 1419  & 18742 $\pm$ 1601\\
              \emph{h$_\mathrm{cut}$} (pc)                 & [500;1500]  &   955 $\pm$  316 &  1070 $\pm$  274\\
              $h_{\rm ra}$ (pc)                 & [1500,3000]  &  2357 $\pm$ 148  &  2436 $\pm$ 125\\
              $h_{\rm rb}$ (pc Gyr$^{-1}$)                 & [700;2700]  &   742 $\pm$ 114  &   743 $\pm$  80\\
              $h_{\rm rc}$ (pc Gyr$^{-2}$)           & [0.20;0,80]  &  0.397$\pm$ 0.210&  0.513$\pm$ 0.191\\
              kp$_{1}$   (pc)                & [3500;6500]  &  3771 $\pm$ 292  &   3798$\pm$  277\\
           $\chi^{2}$      & --------------- & 20.16 $\pm$ 0.12& 20.24 $\pm$  0.14 \\
           $Lr$            & --------------- &  -17957.4       & -18020.0 \\
            BIC            & --------------- &   36055.1       &  36180.3  \\
\noalign{\smallskip} \hline
\end{tabular}
\end{table*}

\begin{table*}[ht]
\caption{Correlations for sim-1.}
\begin{tabular}{cccccccccc}
\hline  & $\gamma_{warp_{+}}$ & $\gamma_{warp_{-}}$ & $R_\mathrm{warp}$ & $\gamma_\mathrm{flare}$ & $R_\mathrm{flare}$ & kp$_{2...7}$ & $R_\mathrm{dis}$ & $h_\mathrm{cut}$ & \\
\hline \noalign{\smallskip}
$\gamma_\mathrm{warp_{+}}$&  1.000 &  0.298 & 0.525 & 0.032 & 0.160&-0.198&  0.172 &-0.192&\\
$\gamma_\mathrm{warp_{-}}$&   0.298 & 1.000 & 0.800 & 0.042&  0.100& -0.287 & 0.080 &-0.321&\\
{\emph{R$_\mathrm{warp}$}}&  0.525 & 0.800&  1.000& -0.334& -0.160& -0.310&  0.246& -0.485&\\
$\gamma_\mathrm{flare}$ &  0.032 & 0.042& -0.334&  1.000&  0.813&  0.178& -0.065&  0.618&\\
{\emph{R$_\mathrm{flare}$}} &   0.160 & 0.100& -0.160&  0.813&  1.000& -0.126&  0.063&  0.339&\\
kp$_{2...7}$ &  -0.198 &-0.287& -0.310&  0.178& -0.126&  1.000& -0.164&  0.467&\\
{\emph{R$_\mathrm{dis}$}} &  0.172 & 0.080&  0.246& -0.065&  0.063& -0.164&  1.000& -0.066&\\
{\emph{h$_\mathrm{cut}$}} & -0.192 &-0.321& -0.485&  0.618&  0.339&  0.467& -0.066&  1.000&\\
\noalign{\smallskip} \hline
\end{tabular} \label{table_corsim1}
\end{table*}

\subsection{Dependence of the warp, flare and scale length with age}

In order to investigate whether the warp and flare parameters
depend on age (as the scale length), and to constrain the warp
formation scenario, four sets of simulations were performed
dividing the ages into two groups, as the number of parameters to
fit would be too high to estimate the warp shape individually for
each Age Class bin. For instance, the first run of optimisation
considered one group for the stars with AC $=1$ and another group
for stars with AC $\geq 2$; the second run, one group for stars
with AC $\leq2$, and the other group for stars with AC $\geq 3$,
and so on. Those tests showed that the dependence on age can be
modelled linearly for most of the parameters, except for the warp
in the second quadrant, which follows roughly a second order
polynomial.

Then, we performed new fits assuming age dependencies of
parameters, either linearly or as a second order polynomial, as
given in Eqs 9 to 15 below with the units provided in Table 6.

\begin{sidewaystable}
\centering \caption{Correlations for sim-2.} \tiny
\begin{tabular}{cccccccccccccccccccc}
\hline  & a$_{\gamma_\mathrm{warp_{+}}}$ & b$_{\gamma_\mathrm{warp_{+}}}$ & c$_{\gamma_\mathrm{warp_{+}}}$ & a$_{\gamma_\mathrm{warp_{-}}}$ & b$_{\gamma_\mathrm{warp_{-}}}$ &a$_{R_{warp}}$ & b$_{R_{warp}}$ & a$_{\theta_{warp}}$ & b$_{\theta_{warp}}$ & a$_{R_{flare}}$ & b$_{R_{flare}}$ & a$_{\gamma_{flare}}$ & b$_{\gamma_{flare}}$ & \emph{R$_\mathrm{dis}$}  & \emph{h$_\mathrm{cut}$} & $h_{\rm ra}$ & $h_{\rm rb}$ & $h_{\rm rc}$ & kp$_{1}$ \\
\hline \noalign{\smallskip}
a$_{\gamma_\mathrm{warp_{+}}}$ & 1.000 &  0.769 &  0.478 &  0.109 &  0.014 &  0.110 &  0.067 &  0.088 &  0.082 & -0.104 &  0.339 &  0.056 &  0.151 & -0.114 & -0.143 & -0.081  & -0.087 & -0.128 &  0.157 \\
b$_{\gamma_\mathrm{warp_{+}}}$ & 0.769 &  1.000 &  0.915 & -0.128 &  0.027 & -0.247 & -0.344 & -0.147 & -0.207 &  0.080 &  0.401 &  0.209 &  0.084 &  0.008 & -0.169 & -0.083  & 0.021  & -0.177 & -0.151 \\
c$_{\gamma_\mathrm{warp_{+}}}$ & 0.478 &  0.915 &  1.000 & -0.224 &  0.090 & -0.287 & -0.441 & -0.165 & -0.289 &  0.151 &  0.369 &  0.219 &  0.058 &  0.044 & -0.162 &  -0.020 & 0.038  & -0.149 & -0.323 \\
a$_{\gamma_\mathrm{warp_{-}}}$ & 0.109 & -0.128 & -0.224 &  1.000 & -0.298 &  0.376 &  0.407 & -0.114 & -0.037 & -0.008 & -0.188 & -0.092 &  0.079 & -0.149 &  0.176 &  0.095  & -0.124 &  0.012 &  0.030 \\
b$_{\gamma_\mathrm{warp_{-}}}$ & 0.014 &  0.027 &  0.090 & -0.298 &  1.000 &  0.096 & -0.113 &  0.173 &  0.566 & -0.061 &  0.423 &  0.151 & -0.035 & -0.025 & -0.004 &  0.446  & -0.300 &  0.220 & -0.097 \\
a$_{R_{warp}}$                 & 0.110 & -0.247 & -0.287 &  0.376 &  0.096 &  1.000 &  0.900 &  0.103 &  0.037 & -0.198 & -0.066 & -0.328 &  0.166 & -0.165 &  0.019 &  0.154  & -0.285 &  0.083 &  0.105 \\
b$_{R_{warp}}$                 & 0.067 & -0.344 & -0.441 &  0.407 & -0.113 &  0.900 &  1.000 &  0.021 & -0.054 & -0.096 & -0.204 & -0.252 &  0.219 & -0.118 &  0.119 &  0.127  & -0.256 &  0.008 &  0.195 \\
a$_{\theta_{warp}}$            & 0.088 & -0.147 & -0.165 & -0.114 &  0.173 &  0.103 &  0.021 &  1.000 &  0.772 & -0.201 &  0.240 & -0.023 &  0.027 & -0.150 & -0.139 &  0.266  & -0.124 &  0.369 &  0.079 \\
b$_{\theta_{warp}}$            & 0.082 & -0.207 & -0.289 & -0.037 &  0.566 &  0.037 & -0.054 &  0.772 &  1.000 & -0.220 &  0.384 &  0.077 & -0.063 & -0.083 & -0.010 &  0.380  & -0.231 &  0.403 &  0.183 \\
a$_{R_{flare}}$                & -0.104 &  0.080 &  0.151 & -0.008 & -0.061 & -0.198 & -0.096 & -0.201 & -0.220 &  1.000 & -0.387 &  0.710 & -0.463 &  0.046 &  0.304 &  0.131  & -0.091 & -0.103 & -0.578 \\
b$_{R_{flare}}$                & 0.339 &  0.401 &  0.369 & -0.188 &  0.423 & -0.066 & -0.204 &  0.240 &  0.384 & -0.387 &  1.000 &  0.190 &  0.350 &  0.041 & -0.211 &  0.193  & -0.006 &  0.212 &  0.004 \\
a$_{\gamma_{flare}}$           & 0.056 &  0.209 &  0.219 & -0.092 &  0.151 & -0.328 & -0.252 & -0.023 &  0.077 &  0.710 &  0.190 &  1.000 & -0.428 &  0.163 &  0.266 &  0.275  & 0.055  & 0.113  & -0.392 \\
b$_{\gamma_{flare}}$           & 0.151 &  0.084 &  0.058 &  0.079 & -0.035 &  0.166 &  0.219 &  0.027 & -0.063 & -0.463 &  0.350 & -0.428 &  1.000 & -0.215 & -0.066 &  0.058  & -0.048 & -0.081 &  0.092 \\
\emph{R$_\mathrm{dis}$}                       & -0.114 &  0.008 &  0.044 & -0.149 & -0.025 & -0.165 & -0.118 & -0.150 & -0.083 &  0.046 &  0.041 &  0.163 & -0.215 &  1.000 & -0.154 &  0.095  & 0.129  & 0.265  & 0.051  \\
\emph{h$_\mathrm{cut}$}                     & -0.143 & -0.169 & -0.162 &  0.176 & -0.004 &  0.019 &  0.119 & -0.139 & -0.010 &  0.304 & -0.211 &  0.266 & -0.066 & -0.154 &  1.000 &  0.095  & -0.009 & -0.029 & -0.032 \\
h$_{\rm ra}$                       & -0.081 & -0.083 & -0.020 &  0.095 &  0.446 &  0.154 &  0.127 &  0.266 &  0.380 &  0.131 &  0.193 &  0.275 &  0.058 &  0.095 &  0.095 &  1.000  & -0.562 &  0.758 & -0.219 \\
h$_{\rm rb}$                       & -0.087 &  0.021 &  0.038 & -0.124 & -0.300 & -0.285 & -0.256 & -0.124 & -0.231 & -0.091 & -0.006 &  0.055 & -0.048 &  0.129 & -0.009 & -0.562  &  1.000 & -0.172 &  0.076 \\
h$_{\rm rc}$                       & -0.128 & -0.177 & -0.149 &  0.012 &  0.220 &  0.083 &  0.008 &  0.369 &  0.403 & -0.103 &  0.212 &  0.113 & -0.081 &  0.265 & -0.029 &  0.758  & -0.172 &  1.000 & -0.062 \\
kp$_{1}$                       & 0.157 & -0.151 & -0.323 &  0.030 & -0.097 &  0.105 &  0.195 &  0.079 &  0.183 & -0.578 &  0.004 & -0.392 &  0.092 &  0.051 & -0.032 & -0.219  &  0.076 & -0.062 &  1.000 \\
\noalign{\smallskip} \hline
\end{tabular} \label{table_corsim2}
\end{sidewaystable}

\bigskip
\begin{equation} \label{equ:warp_age}
\displaystyle \gamma_\mathrm{warp_{+}} =
a_{\gamma_\mathrm{warp_{+}}} - b_{\gamma_\mathrm{warp_{+}}} \times
<\rm age_{i}> + c_{\gamma_\mathrm{warp_{+}}}\times <\rm age_{i}>^{2} ,
\end{equation}

\begin{equation} \label{equ:warp_age2}
\displaystyle \gamma_\mathrm{warp_{-}} =
a_{\gamma_\mathrm{warp_{-}}} + b_{\gamma_\mathrm{warp_{-}}} \times
<\rm age_{i}>,
\end{equation}

\begin{equation} \label{equ:rwarp_age}
\displaystyle R_\mathrm{warp} = a_{R_\mathrm{warp}} -
b_{R_\mathrm{warp}}\times <\rm age_{i}>,
\end{equation}

\begin{equation} \label{equ:thetawarp_age}
\displaystyle \theta_\mathrm{warp} = a_{\theta_\mathrm{warp}} -
b_{\theta_\mathrm{warp}} \times <\rm age_{i}>,
\end{equation}

\begin{equation} \label{equ:rflare_age}
\displaystyle R_\mathrm{flare} = a_{R_\mathrm{flare}} +
b_{R_\mathrm{flare}} \times <\rm age_{i}>,
\end{equation}

\begin{equation} \label{equ:gflare_age}
\displaystyle \gamma_\mathrm{flare} = a_{\gamma_\mathrm{flare}} +
b_\mathrm{flare} \times <\rm age_{i}>,
\end{equation}

\bigskip

The scale length also has a tendency to decrease with age in those
simulations. Hence, we adopted a dependence as shown in Eq. 15,
which adequately follows the variations seen in the tests with two
age groups.

\bigskip

\begin{equation} \label{hr}
\displaystyle h_{\rm R} = h_{\rm ra} + h_{\rm rb} \times
{\rm exp}(-h_{\rm rc}<\rm age_{i}>),
\end{equation}

\bigskip

\noindent where \emph{$<\rm age_{i}>$} is the mean age of Age
Class $i$, as indicated in Table 1.

\begin{figure*}[ht]
\centering
\resizebox{8.2cm}{!}{\includegraphics[angle=90,scale=.90]{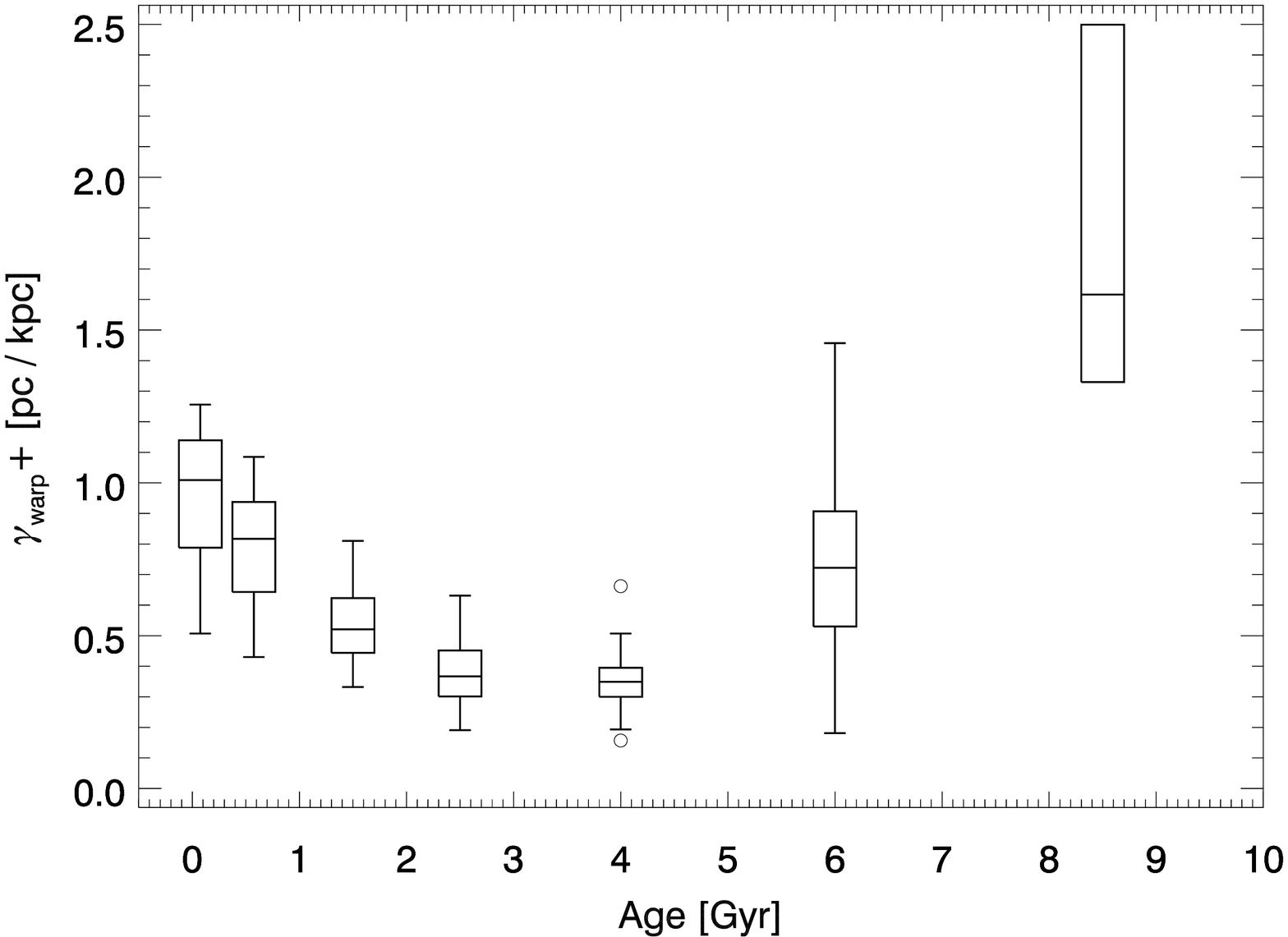}}
\resizebox{8.2cm}{!}{\includegraphics[angle=90,scale=.90]{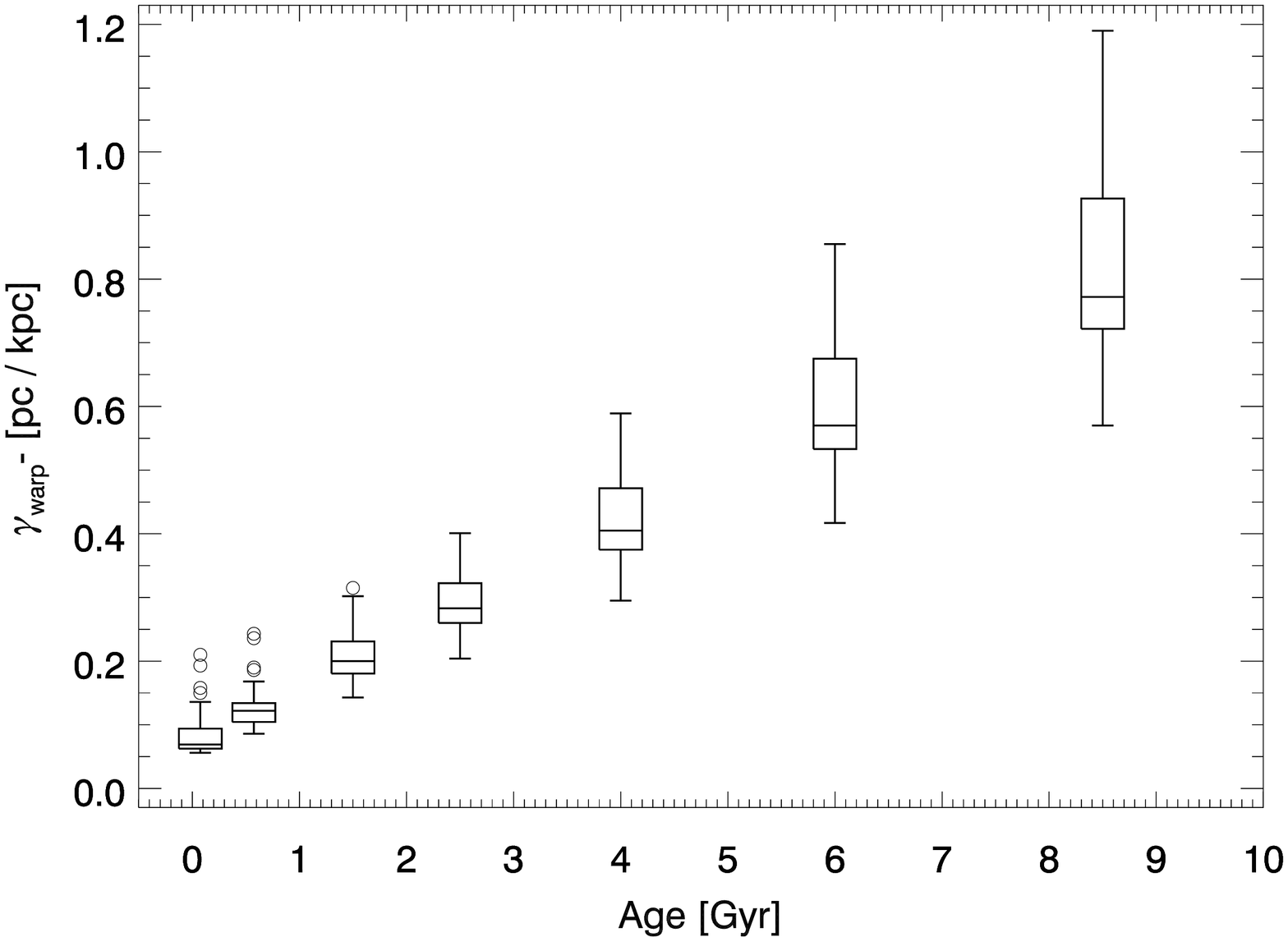}}
\resizebox{8.2cm}{!}{\includegraphics[angle=90,scale=.90]{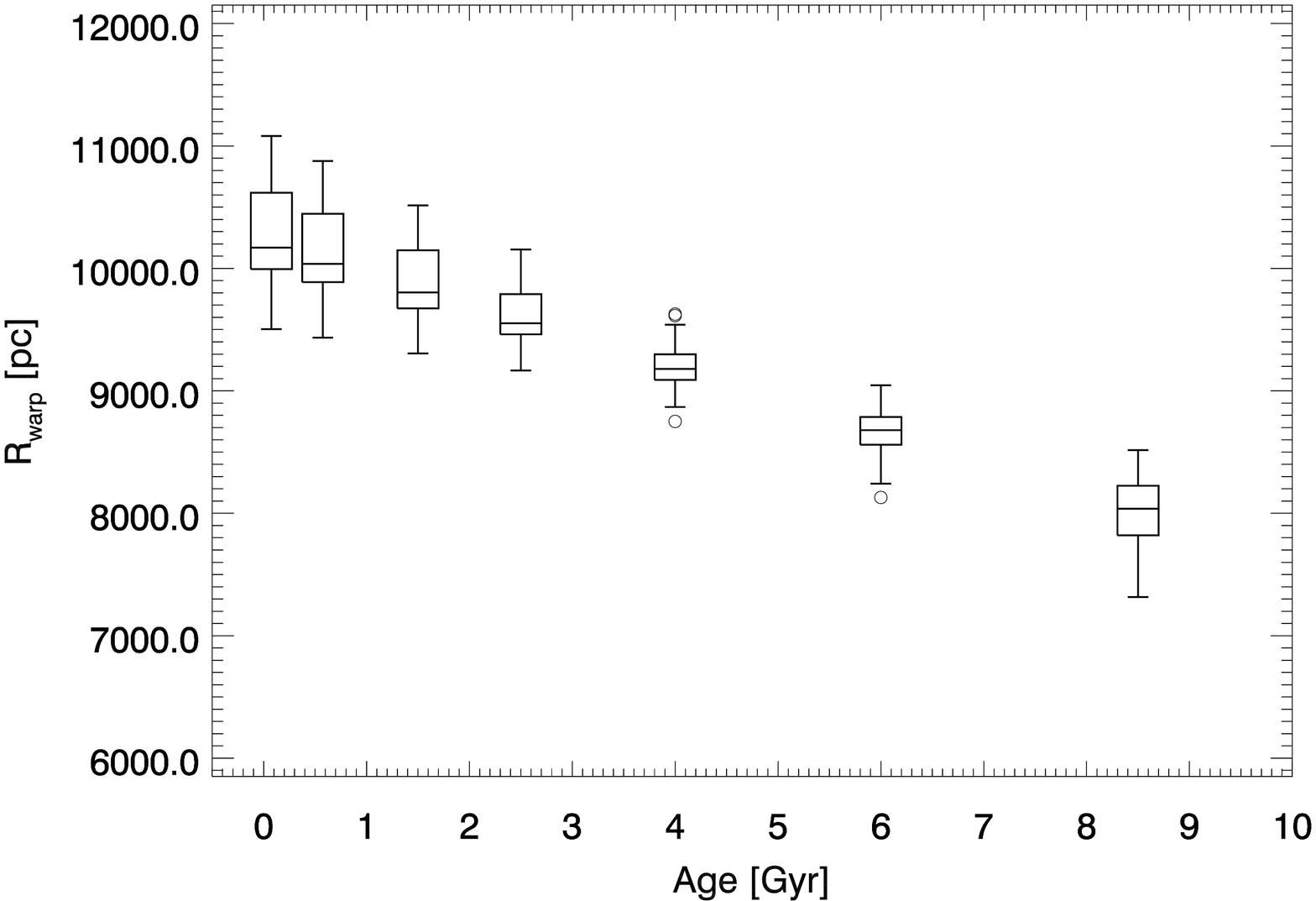}}
\resizebox{8.2cm}{!}{\includegraphics[angle=90,scale=.90]{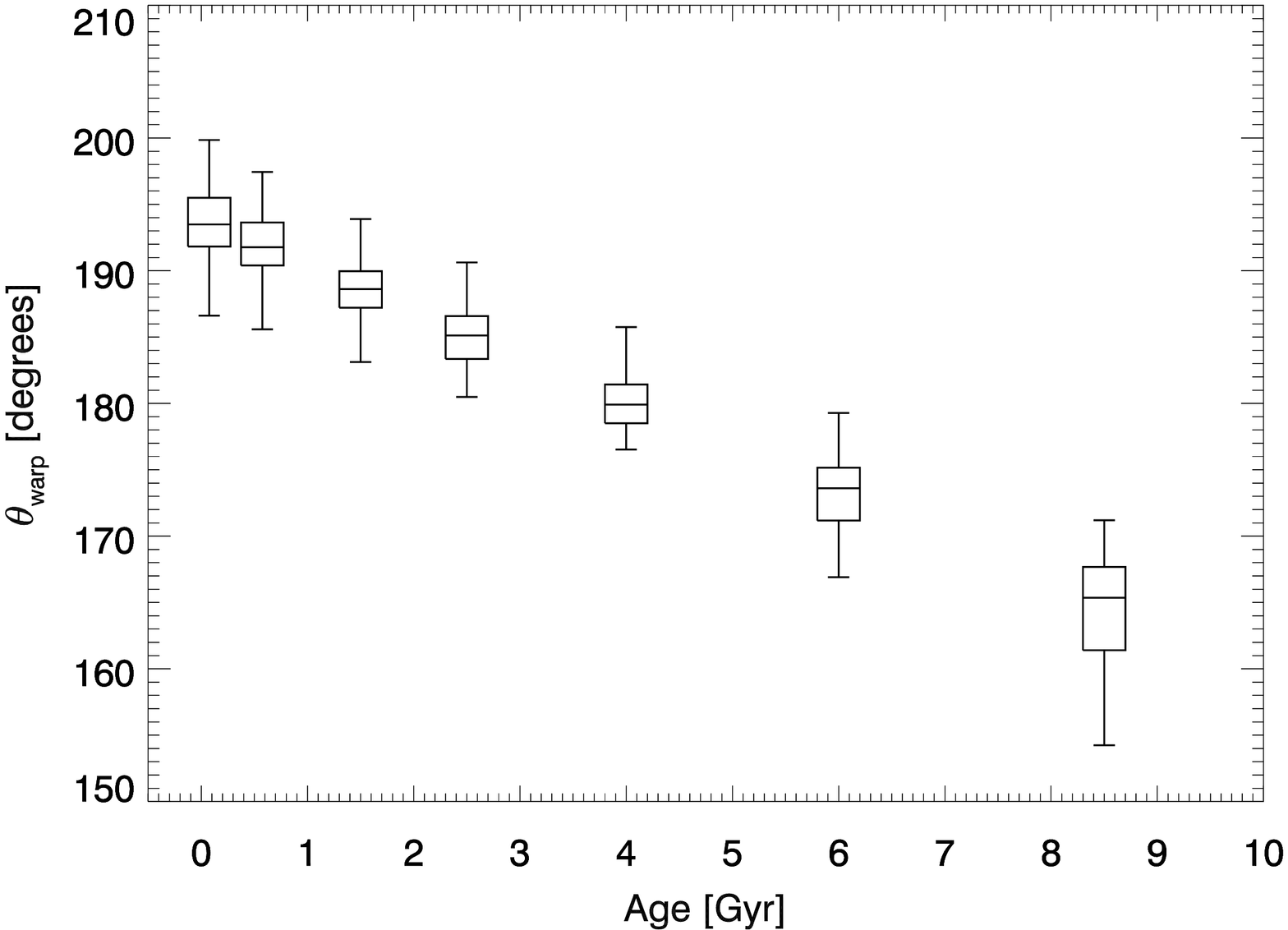}}
\caption{Dependence of warp parameters: amplitude in the second
quadrant (top left) and in the third quadrant (top right);
starting radius (bottom left) and angle (bottom right) obtained in
sim-2 using Eqn 9 to 12 with parameters provided in Table 6.
Circles represents outliers (see text). In the top left panel, the
whiskers are not shown at 8.5 Gyr because they are too large
(0.371 and 3.639).} \label{warpage}
\end{figure*}

\begin{figure}[h]
\includegraphics[scale=0.34,angle=90]{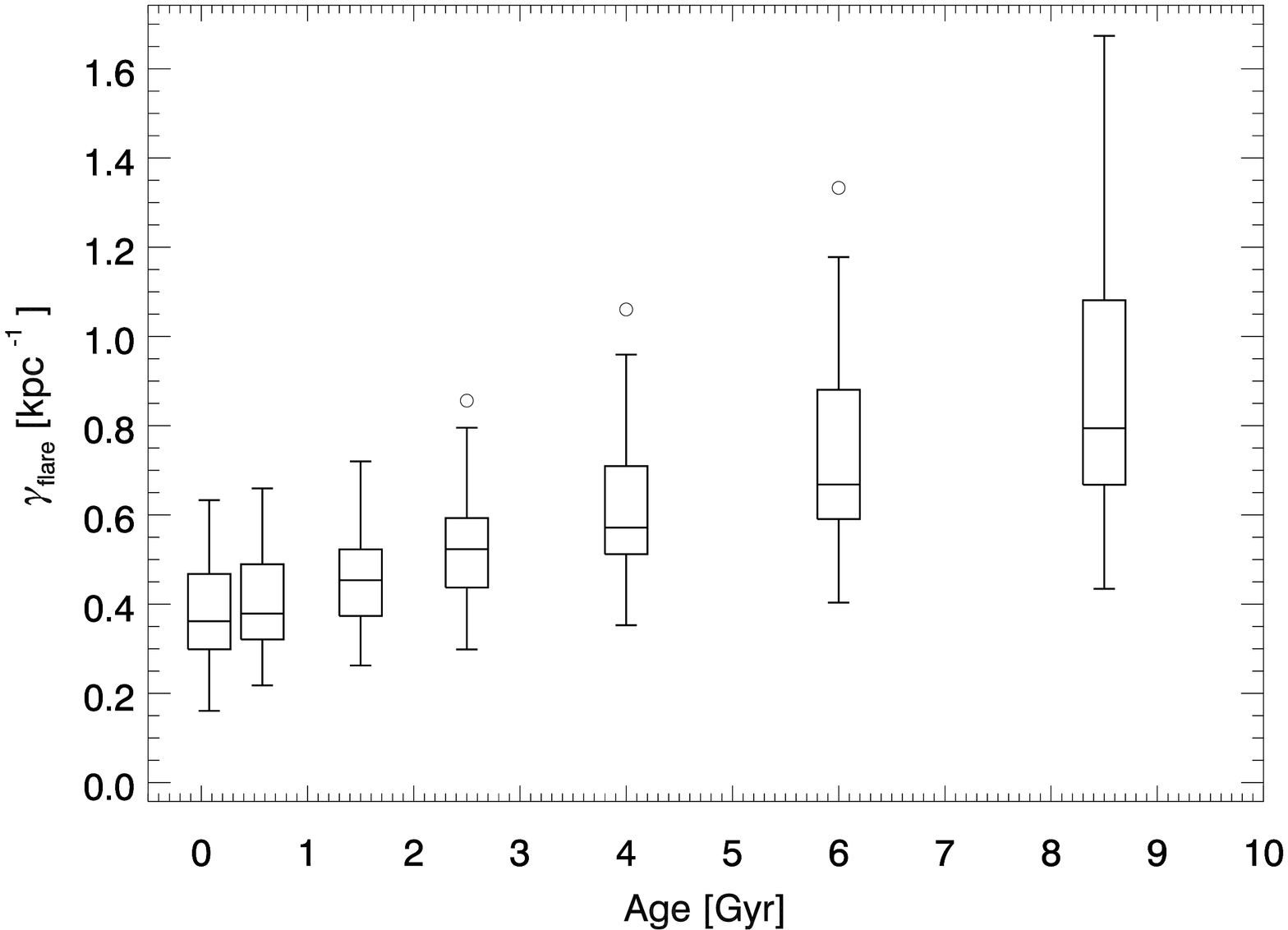}
\includegraphics[scale=0.34,angle=90]{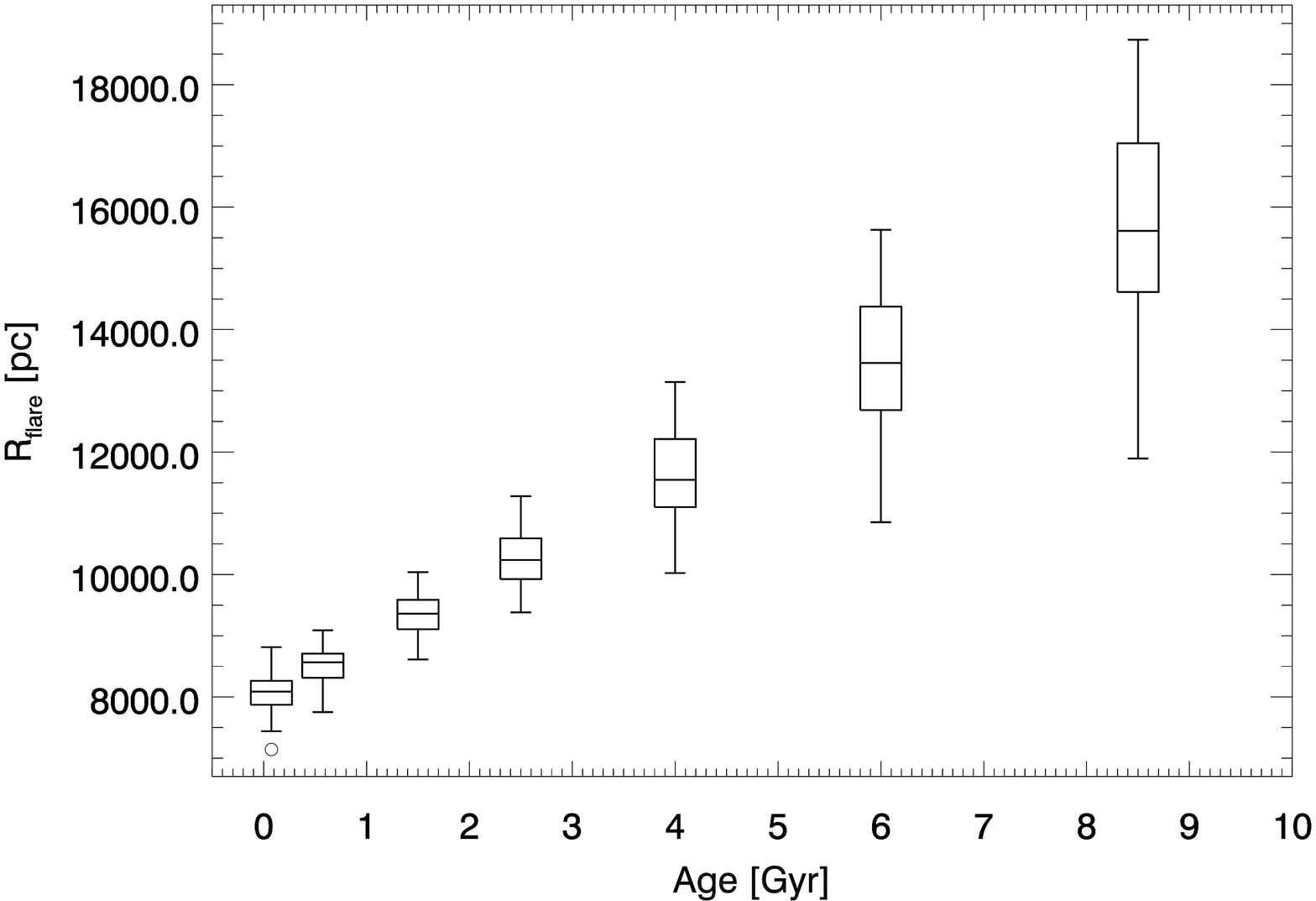}
\centering \caption{Dependence  with age of the flare parameters,
as given in Table 6 and Eqn 13 and 14: amplitude (upper panel) and
starting radius (lower panel).} \label{flareage}
\end{figure}

Using the equations above, with the range of parameters given in
Table 6 (second column), we have performed a new set of
optimisations (called sim-2) with 100 independent runs, using the
ranges of parameters presented in Table~6, together with the best
fit values and estimated uncertainties. Eqn 15 is applied on ages
from 0.15 to 10 Gyr, while the first age bin is fitted with the
parameter kp$_{1}$, as in sim-1. The standard errors can be high
in some cases, such as $c_{\gamma_\mathrm{warp_{+}}}$ and
$b_{\gamma_\mathrm{flare}}$. Although their impact on our
estimates of the warp and flare shapes is limited and concerns
mainly oldest stars, as will be shown in Figs.~\ref{warpage} and
\ref{flareage}.

As can be seen in the comparison of $\chi^{2}$ in tables 5 and 6,
there is an improvement of $\chi^{2}$ in sim-2 compared to sim-1.
To estimate the number of parameters and the effective improvement
in the $\chi^{2}$, we compute the Bayesian information criterion
(BIC) \citep{Schwarz1978}. We follow the recipes of
\cite{Robin2014} first computing the reduced likelihood ($Lr$,
Eqn~\ref{likelihood}) for a binomial statistics
\citep{Bienayme1987a}, as described below:

\begin{equation}
Lr= \sum_{i=1}^{n} q_i \times (1 -R_i + \mathrm{ln}(R_i)),
\label{likelihood}
\end{equation}

\noindent where $f_i$ and $q_i$ are the number of stars in bin $i$
in the model and the data, and $R_i=\frac{f_i}{q_i}$.

The BIC is then computed following the formula from
\cite{Schwarz1978}, $BIC = -2\times Lr + k\times \mathrm{ln}(n)$.
It penalises models with a larger number of parameters and allows
to compare the goodness of fit of different models with a
different number of fitted parameters. $n$ is the number of bins
on which the data are fitted and \emph{k} the number of fitted
parameters. In our case, $n$ = 1615, and \emph{k} = 10 and 19 for
sim-1 and sim-2, respectively.

Then, $Lr$ (Eqn~\ref{likelihood}) and the BIC were computed for
the three fitting procedures with their values showed in Table 5
(sim-1) and Table 6 (sim-2 and sim-3). The BIC for sim-2 and sim-3
are quite similar despite the fact that sim-3 incorporates extra
parameters to model the ripples, but we did not adjust these
parameters. Hence the number of fitted parameters \emph{k} is the
same in sim-2 and sim-3. The comparison between sim-1 and sim-2
highlights the fact that sim-2, despite using more parameters than
sim-1, is preferred on the criterion of goodness of fit.

\begin{figure*}[h]
\centering
\includegraphics[height=14.0cm,angle=90.0]{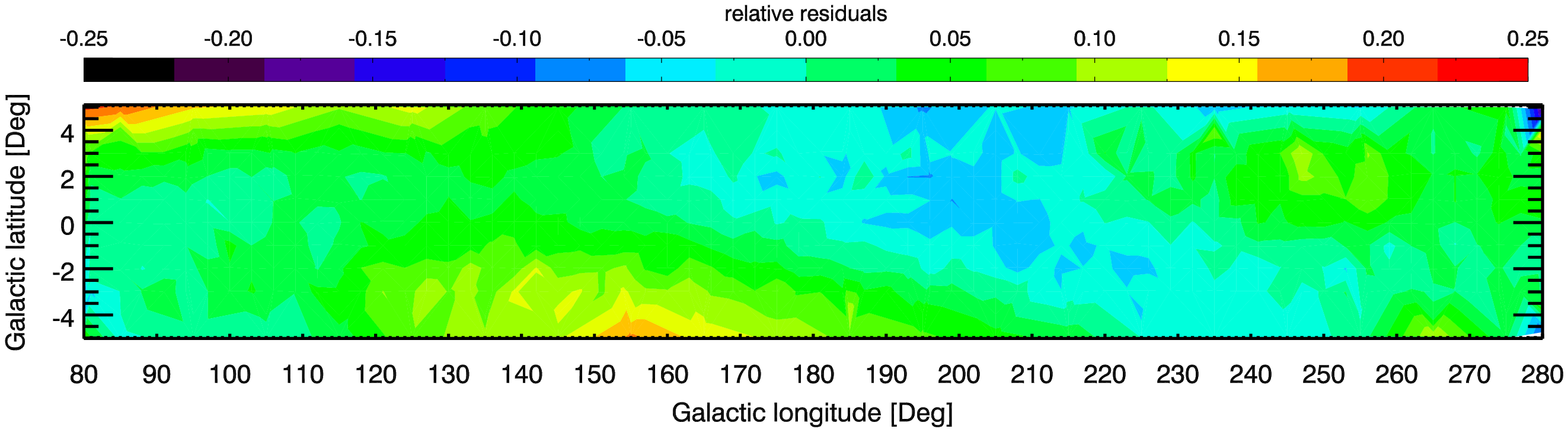}
\caption{Map of the relative differences between sim-2 and sim-1
(\emph{N$_{\rm sim-2}$}-{\emph{N$_{\rm sim-1}$}})/{\emph{N$_{\rm
sim-1}$}}.} \label{maps_sim2msim1}
\end{figure*}

The values obtained for each parameter in sim-2 are given in Table
6 (third column). The map of the relative differences between
sim-1 and sim-2 (\emph{N$_{\rm sim-2}$}-\emph{N$_{\rm
sim-1}$})/\emph{N$_{\rm sim-1}$} is presented in
Figure~\ref{maps_sim2msim1}. The average difference is around
10\%. There are two main large  area where the two fitted models
differ. The first one at $\ell \sim 150^{\circ}$ and \emph{b} $<
-2^{\circ}$ where sim-2 produces more counts than sim-1, and the
area around $\ell \sim 200^{\circ}$ where it produces less. The
shape of the residuals seems to indicate that the change in the
mean scale length with age in sim-2 affects the counts in the
anti-centre, as expected. But also the shape of the warp varying
with time has an impact on the star counts at the level of about
10 to 15\% in the specific area at $\ell \sim 150^{\circ}$ and
\emph{b} $< -2^{\circ}$.

\begin{figure}
\centering
\includegraphics[scale=0.40,angle=90]{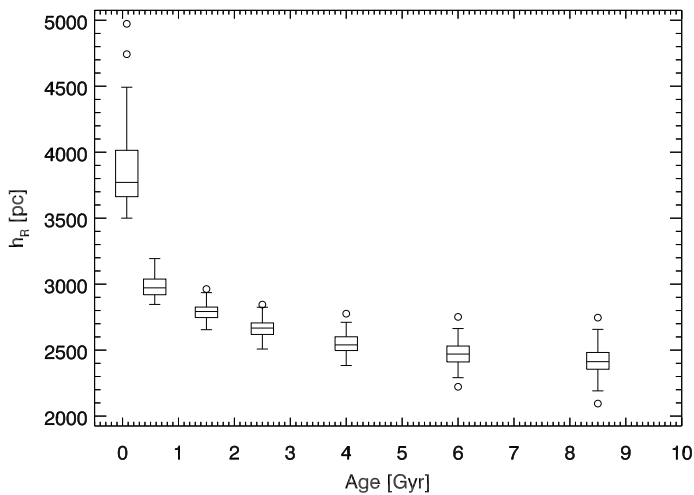}
\centering \caption{Scale length as a function of age (see text
and Eqn 15).} \label{hRage}
\end{figure}

The variation of the warp, flare and scale length with age over
the 100 different independent solutions are presented in
Figs~\ref{warpage} to \ref{hRage}, respectively. The box plots
were produced by using IDL Coyote's
Library\footnote{http://www.idlcoyote.com/documents/programs.php};
the box encloses the InterQuartile Range (IQR), defined as
\emph{Q$_{3}$} - \emph{Q$_{1}$}, in which \emph{Q$_{3}$} and
\emph{Q$_{1}$} are the upper and lower quartile, respectively. The
whiskers extend out to the maximum or minimum value of 100
independent solutions, or to the 1.5 times either the
\emph{Q$_{3}$} or \emph{Q$_{1}$}, if there is data beyond this
range. The small circles are outliers, defined as values either
greater or lower than 1.5 $\times$ \emph{Q$_{3}$} or
\emph{Q$_{1}$}.

We note that all warp parameters significantly change with age
(Figure~\ref{warpage}). The slope of the warp for both second and
third quadrants varies as a function of age. The shapes
significantly vary from second to third quadrant, even though
error bars are larger in the third quadrant and for stars older
than 6 Gyr. One common aspect resides on the fact that the
amplitude increases for stars with age greater than 4 Gyr for both
quadrants.

These figures show that the warp shape changes significantly with
time (or with the age of the tracers), and the variations are
different in the second and in the third quadrants. The starting
radius of the warp changes monotonically with age, as well as the
node angle. But for young stars, the slope of the warp is very
different between second and third quadrants. This confirms the
strong dissymmetry between the region of the plane moved to
positive latitudes (first and second quadrants) and the opposite
region (third and fourth quadrant), but points out that this
dissymmetry is mainly due to young stars, while old stars follow a
more symmetrical warp, even though the error bars for old stars
are larger.

The asymmetry of the warp was long identified in HI gas
\citep{Burton1988,Levine2006,Kalberla2008} and even in stars
\citep{LC2002,Reyle2009}. But this is the first time that it is
shown that the shape varies with the age of the tracers
considered.

The flare also exhibits a time dependence (Figure~\ref{flareage}).
Despite the fact that error bars of the flare amplitude increase
with time, there is a clear trend showing that the amplitude
increases with age. The starting radius of the flare also varies
with age. We shall consider in Section 5.4 whether these features
can be due to correlations.

Figure~\ref{hRage} shows the scale length variation with age
obtained with sim-2 using Equation (15). The values of the fitted
parameters are given in Table 6. As can be seen in this figure,
the scale length clearly decreases with age. This result, which is
related to the thin disc formation scenario, will be discussed in
Section 6.

\subsection{Disc truncation}

As seen in Table~5 with sim-1, the disc truncation
(\emph{R$_\mathrm{dis}$}) is found at approximately 16.1 $\pm$ 1.3
kpc. The scale (\emph{h$_\mathrm{cut}$}) of cutoff is equal to
0.72 $\pm$ 0.27 kpc. It is such that the density drops by 90\%
within 1 kpc, therefore, the cutoff is sharp. In sim-2, when age
dependence for warp, flare and scale length is considered the disc
truncation obtained is 19.4 $\pm$ 1.4 kpc and considering the
ripples (sim-3, see Section 5.5) is 18.7 $\pm$ 1.6 kpc, still in
agreement at the $1-\sigma$ level with its value in sim-1.

We note that the disc truncation is found at a larger
Galactocentric distance than in the standard model, where it was
assumed 14 kpc. The determination is here more robust than in
\cite{Robin1992a} from which the standard model was based, since
only one direction was used for that determination.

\subsection{Parameter correlations}

In order to investigate whether the parameters are correlated, we
computed the correlation coefficients in the optimisation
procedure sim-1, shown in Table~\ref{table_corsim1}, and in
Figure~\ref{AppendixSim1} in the Appendix.
\emph{R$_\mathrm{flare}$} and $\gamma_\mathrm{flare}$ are fairly
correlated because we have considered only low latitude fields in
our study. The same problem was mentioned by \cite{LC2002} and
\cite{LC2014}. A significant correlation also appears between
\emph{R$_\mathrm{warp}$} and the warp amplitude specially in the
third quadrant.

Table~\ref{table_corsim2} shows the correlations among the
parameters for sim-2. As can be seen, some parameters are
correlated (notably the starting radius and the amplitudes of the
warp and flare). This is expected because the photometric
distances are not precise enough. Hence, the exact position where
the flare starts is not well determined, and it probably starts
more smoothly than a linear increase starting at a given point.
But the overall measurements of the warp and flare dependencies
with age are clear enough and not impacted by those correlations.
To avoid these correlations, in the future several complementary
studies should be considered. Firstly, looking at higher
latitudes, covering latitudes from -30\deg~to +30\deg~in the
anti-centre direction would allow disentangling the effect of the
flare from the radial truncation. Secondly, the correlation
between the starting radius and the amplitude is due to a
simplistic linear shape used. Using samples with good distance
estimates, such as Gaia data or red clump giants, would allow to
determine better the shape of the starting radius and the
amplitude of the flare.

\subsection{Is there an improvement considering the ripples ?}

We also considered in another set of simulations (sim-3), the
contribution of ripples, an oscillation in the number of star
counts, as described by \cite{Xu2015} without fitting any
parameter of the ripple itself. They have used SDSS data towards
110$^\circ$~$<$ $\ell$~$\le$~229$^\circ$ in order to study the
structures in the outer Galaxy. We have used the same equations
for the northern and southern, as provided by \cite{Xu2015} in
Section 6 of their paper.

Basically, when computing \emph{z$_{\rm w}$} we add  the
\emph{z$_{\rm wripple}$} given by \cite{Xu2015}. The resulting
parameters and the $\chi^{2}$ of the solution are given in Table
6, showing no improvements compared with sim-2 but rather a
slightly higher BIC value.

\section{Discussion}

\subsection{Thin disc scale length}

The disc scale length is the most important unknown in
disentangling the contributions from the disc and the dark halo to
the mass distribution of the Galaxy \citep{Dehnen1998, Bovy2013}.
Recently, the thin disc scale length has been studied from
photometric and spectroscopic large scale surveys. For example,
\cite{bovy2012} determine various scale lengths for mono-abundance
populations. Assuming that the thin disc is defined by the low
$\alpha$ abundance population, they show that the scale length can
depend on the metallicity within the thin disc. In their sample
high metallicity stars have a shorter scale length than low
metallicity ones, with noticeably large error bars, which is in
contradiction with our work if metallicity is anti-correlated with
age. However the mean orbital Galactic radii of the low
metallicity stars are much larger, which implies that their sample
of low metallicity stars comes from the outer Milky Way, while the
high metallicity sample comes from the inner Milky Way. In the
end, the sample from the outer Milky Way has larger scale length,
as expected from the inside-out scenario. The scale lengths of the
mono-abundance populations of the thin disc range from 2.4 to 4.4
kpc, while we found a very similar scale length range of 2.3 to
3.9 kpc.

Our result of a short scale length for the main thin disc (as
stars older than 3 Gyr dominates the counts in the Milky Way) is
in general agreement with previous results. Table~\ref{table_hR}
shows most of the scale length determinations from different
authors during the last twenty years as well as the tracer used
and field coverage. It is noticeable that most studies using IR
data found a short scale length for the thin disc, compared with
longer scale length obtained from visible data. This is
understandable if the IR tracers are in the mean older than the
visible tracers (young stars emit mostly in the visible). However,
this could also be due to the extinction which perturbs the
analysis in the visible.

\citet{LC2002} constrained the outer disc (scale length, warp and
flare) from a study of 2MASS data in 12 fields but only four
different longitudes (155\deg, 165\deg, 180\deg~and 220\deg),
which does not include fields where the warp is stronger (around
90\deg~and 270\deg). They only use a selection of red giant stars
from CMDs and they apply a simple modelling approach assuming a
luminosity function and density law for the disc. They find a thin
disc scale length of 1.91$^{0.20}_{-0.16}$ kpc, which is in
excellent agreement with our determination for stars older than 3
Gyr, but smaller than our mean scale length. Their sample is
dominated by red giants. Hence it does not include higher mass and
younger stars, which might explain the difference.

\cite{Yusifov2004} studied the distribution of pulsars from the
\cite{Manchester2005} catalogue. Although \cite{Momany2006} claims
that the catalogue is incomplete in the outer Galaxy, their scale
length for these young objects is 3.8 $\pm$ 0.4 kpc, in close
agreement with our young star sample.

As pointed above, \cite{Kalberla2008} by fitting HI distribution
obtained a radial exponential scale length of 3.75 kpc in the
mid-plane. Even if they do not give an estimate of their error, it
is in excellent agreement with our very young star scale length of
3.90 $\pm$ 0.28 kpc.

For the first time, we report a clear evolution of the scale
length with time among field stars. There have been previous
evidences of young stellar associations and young open clusters at
distances where the old stars seem to be no more present
\citep{Robin1992b}. However, it was long known that the HI disc
has a longer scale length than the stars in the mean. Hence, it is
not too surprising to see that very young stars have a scale
length similar to the gas.

\subsection{Disc truncation}

The disc truncation has been often observed in external galaxies. The
existence of this cut-off can be related to a threshold in the
star formation efficiency, when the gas density drops under
certain value \citep{Kennicutt1989,vanderkruit1979}.

We found a radius for disc truncation equal to
\emph{R$_\mathrm{dis}$} = 16.1 $\pm$ 1.3 kpc in the case of sim-1.
This value increases up to 18 kpc in the case of sim-2 when the
scale length is assumed to vary with time. We have not tested a
possible dependence of the truncation on age, in order to avoid
increasing too much the number of free parameters. We shall
consider this point in the future by extending the fit to higher
latitudes, in order to decrease the degeneracy between the flare
and the truncation.

\cite{Habing1988} from OH/IR stars and \cite{Robin1992a} from UBV
photometry found the edge of Galactic disc located at 14-15 kpc
and 14 kpc, respectively. \cite{Ruphy1996} using DENIS data similarly found
 15 $\pm$ 2 kpc for the disc truncation distance. These values are in good agreement with our determination.
Using photometry for IPHAS stars towards 160$^\circ \le \ell \le
200^\circ$ ($|b| \le 1.0^\circ$), \cite{Sale2010} obtained a
truncation radius of 13.0 $\pm$ 1.1 kpc, slightly smaller than
ours, but still within the uncertainties. \cite{LC2002} did not
find a disc truncation at distances closer than 20 kpc, but
indicated that they could not distinguish between a truncation or
a flare.

\begin{table*}[t]
\caption{Scale length determinations from other authors with
tracers and field coverage.}
\begin{tabular}{cccc}
\hline  Author & Tracer & Coverage & Scale length (kpc) \\
\hline \noalign{\smallskip}
\cite{Robin1992a, Robin1992b}  & optical star counts & $\ell=189^\circ$ & 2.5 $\pm 0.3$  \\
\cite{Ruphy1996}     & DENIS  & $\ell= 217, 239^\circ$ &2.3 $\pm 0.1$   \\
\cite{Porcel1998}    & TMGS (NIR star counts) &  30$^\circ < \ell < 70^\circ$ for $|b| >$ 5$^\circ$ & 2.1 $\pm 0.3$    \\
\cite{Freud1998}    & COBE & whole sky & 2.2  \\
\cite{Ojha2001}     & 2MASS & 7 fields for $|b| >$ 12$^\circ$ &  2.8 $\pm$ 0.3  \\
\cite{Chen2001}    & SDSS & 279 deg$^{2}$ at high-latitude (49$^\circ$ $< |b| <$ 64$^\circ$ & 2.25   \\
\cite{Siegel2002}  & optical star counts & optical star counts 14.9 deg$^{2}$ at $|b| >$ 25$^\circ$ & 2-2.5    \\
\cite{LC2002}      &  2MASS & $\ell= 180^\circ, 220^\circ$ (b$=0^\circ,3^\circ,6^\circ,9^\circ,12^\circ$)   & 1.97 $^{+0.15}_{0.12}$   \\
                            & & and $\ell= 155^\circ, 165^\circ$ ($b=0^\circ$)  &    \\
\cite{Larsen2003}    & APS catalog (optical) & 16 deg$^{2}$ for $|b| >$ 20$^\circ$ & 3.5    \\
\cite{Cabrera-Lavers2005}  & 2MASS & $|b| >$ 25$^\circ$  &  2.1   \\
\cite{Bilir2006}  & SDSS & 6 fields (41$^\circ \le b \le 52^\circ$) &  1.9   \\
\cite{Karaali2007}  & SDSS & $\ell = 60^\circ, 90^\circ, 180^\circ$ ($b = 45^\circ$, 50$^\circ$) &  1.65-2.52   \\
\cite{Cignoni2008} & optical open clusters & NGC6819, NGC7789, NGC2099  &   2.24-3.00  \\
\cite{Juric2008}    & SDSS & $|b| >$ 25$^\circ$ (6500 deg$^{2}$) & 2.6 $\pm$ 0.5   \\
\cite{Yaz2010}    & SDSS & 22 fields (0$^\circ < \ell \le 260^\circ$) for  (44.3$^\circ \le b \le 45.7^\circ$)  & 1.-1.68   \\
\cite{Chang2011}  & 2MASS & $|b| >$ 30$^\circ$ & 3.7 $\pm$ 1.0   \\
\cite{McMillan2011} & kinematic  & ------------ & 3.00 $\pm$ 0.22    \\
\cite{bovy2012}    & SDSS/SEGUE & 28,000 G-type dwarfs ($|b| >$ 45$^\circ$) & 3.5 $\pm$ 0.2   \\
\cite{Robin2012}   & 2MASS & $|\ell| \le 20^\circ$ (-10$^\circ \le b \le 10^\circ$) & 2.5   \\
\cite{Cheng2012} & SDSS/SEGUE & 50$^\circ < \ell \le 203^\circ$ ($b=-15.0^\circ$,-12.0$^\circ$, $\pm$ 8$^\circ$,10.5$^\circ$,14$^\circ$,16$^\circ$) & 3.4$^{+2.8}_{-0.9}$   \\
\noalign{\smallskip} \hline
\end{tabular} \label{table_hR}
\end{table*}
\bigskip

\subsection{Warp parameters and origin}

Several scenarii have been proposed to explain the Galactic warps,
the most cited being a dark halo-disc misalignment and possible
interactions with nearby galaxies (or flyby galaxies).
\cite{Perryman2014} stated that the misalignment of the disc
inside the dark halo might produce changes in the warp angle with
time. But in this case, all components of the Galaxy should be
perturbed independently of their age. The warp angle dependence
that we see could be consistent with this scenario if the time
dependence reflects a precession. Alternatively,
\cite{Weinberg2006} produced dynamical simulations of the flyby of
the Magellanic Clouds around the Galaxy which could produce a
strongly asymmetric warp varying with time.

However, \cite{Reshetnikov2016} analysed the global structure of
13 edge-on spiral galaxies using SDSS data. They pointed out that
the warps found in those galaxies are generally slightly
asymmetrical. They studied the relation of the strength and
asymmetry of the warps with the dark halo mass. They showed that
galaxies with massive halos have weaker and more symmetric warps,
concluding that these dark halos play an important role in
preventing strong and asymmetric warps. In the case of our Galaxy,
the M$_{\rm tot}/M_{*}$ amount to about 10 \citep{Robin2003} and
the warp angle (computed as seen in edge-on galaxies) is less than
1\deg. Typical galaxies in Reshetnikov's sample with this mass
ratio have warp angles less than 10\deg~and an asymmetry of less
than 5\deg. Hence, our Galaxy compares well with these edge-on
galaxies having a relatively large dynamical to stellar mass ratio
and a weak warp, even though it is asymmetrical.

The Galactic warp can be observed in different components, such as
dust, gas and stellar. \cite{Reyle2009} using BGM and 2MASS
described the warp and flare, comparing their determination with
those provided by several authors from different tracers, such as
dust, gas and stars. In this work, they considered
\emph{R$_\mathrm{warp}$} = 8.4 kpc, $\gamma_\mathrm{warp}=$ 0.09
pc kpc$^{-1}$ for a scale length equal to 2.2 kpc, the same value
obtained by \cite{Derriere2001}. Their starting radius was a bit
shorter than our mean value of 9.18 kpc.

\begin{figure}[h]
\includegraphics[scale=0.39,angle=0]{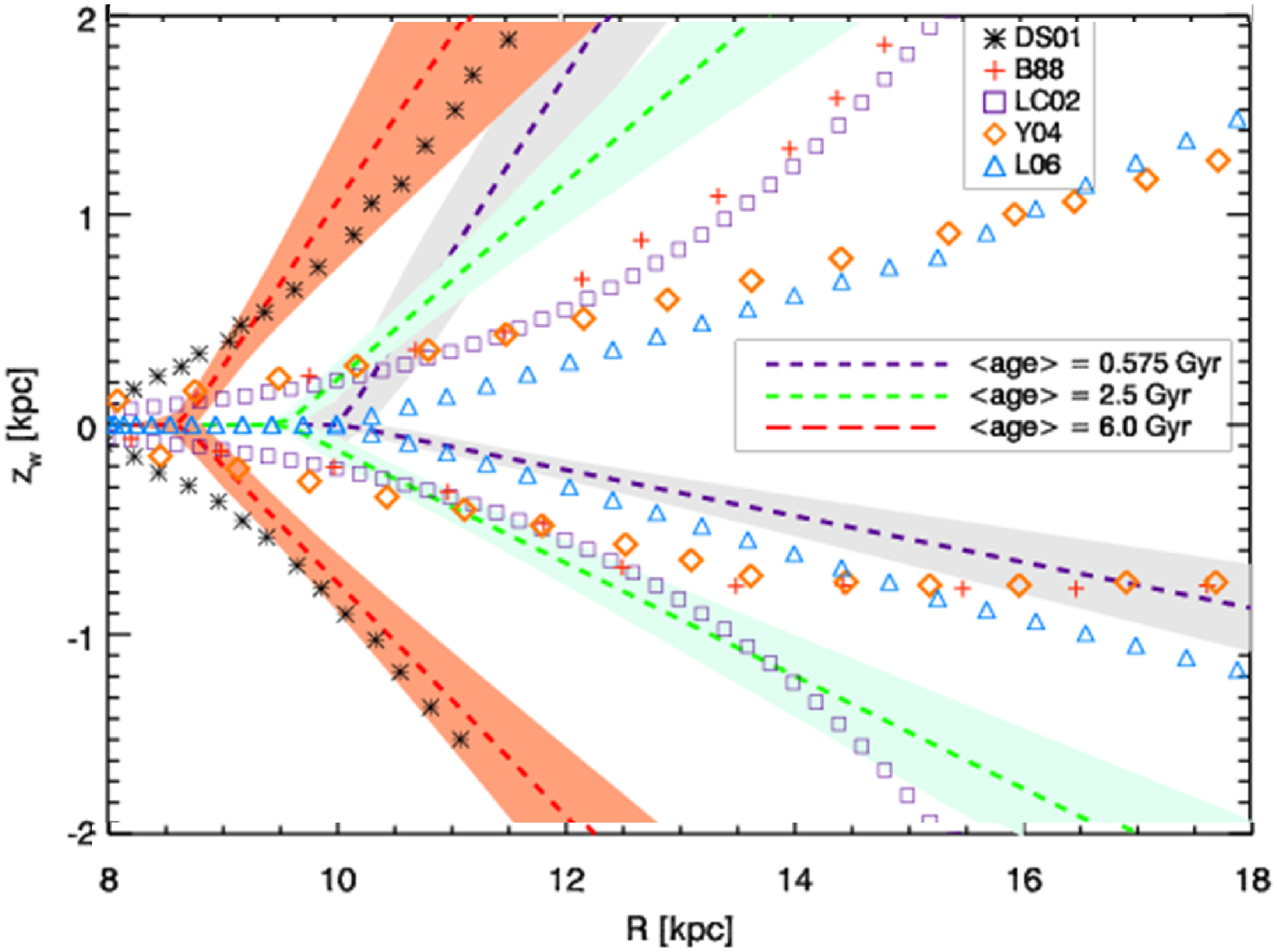}
\centering \caption{Comparison of the maximum height of the warp
as a function of Galactocentric distance obtained in sim-2 for
three different mean ages (dashed lines) with other authors
(symbols): \cite{LC2002} from red giants (LC02: black open
squares), \cite{Yusifov2004} from pulsars (Y04: orange diamonds),
\cite{Burton1988} from HI gas (B88: orange plus signs),
\cite{Levine2006} model fitted on HI (L06: blue open triangles),
\cite{DS2001} from COBE/DIRBE data (DS01: black asterisks).
Positive (resp. negative) values of \emph{z$_{\rm w}$} refer to
second (resp. third) quadrant.} \label{warpwithR}
\end{figure}

They found evidence that the warp is asymmetric but they were not
able to find a good S-shape for the warp in the third quadrant. In
the present work, we have adjusted distinct values for the slope
of the warp at second and third quadrants. We found a value larger
than the one obtained by \cite{Reyle2009}, with a slope in the
second quadrant approximately three times larger than in the third
quadrant. Analysing dust extinction distributions,
\cite{Marshall2006} also determined that the warp in the second
quadrant has a larger amplitude  than in the third quadrant,
$\gamma_\mathrm{warp}=$ 0.14 and 0.11 pc kpc$^{-1}$, respectively.
They also pointed out that the warp seems to start earlier in the
third quadrant. However, the two parameters
(\emph{$R_\mathrm{warp}$} and $\gamma_\mathrm{warp}$) are not
completely independent when the warp model is assumed linear.

The difference can be well explained by the fact that we have
varied other parameters which were not considered by these
authors. They also did not consider the dependence with age, and
in the present study we considered a larger range of longitudes,
latitudes and limiting magnitude to constrain the model.

\cite{DS2001} modelled the Galactic structure by using COBE/DIRBE
data at near and far-infrared assuming a warp with a quadratic
function
$z_\mathrm{warp}(\emph{R})=27.4(\emph{R}-R_\mathrm{warp})^2\sin(\phi)$,
with \emph{R$_\mathrm{warp}$} = 7 kpc. \cite{LC2002} model the
warp with a different shape than ours. The amplitude of their warp
is $2.1\times 10^{-19}\times \emph{R}^{5.25}$. It means that the
mid-plane is going up to 1.2 kpc at \emph{R} = 14 kpc, their node
angle is found to be -5\deg $\pm$ 5\deg. They are less sensitive
to very young stars than we are because they select mainly red
giants in their sample.

\cite{Yusifov2004} models the warp that pulsars follow and find a
warp node angle of 14.5\deg, at $1-\sigma$ of our angle for
youngest stars. \cite{Levine2006} present a very complex warp
shape in HI, which extends at much larger distances, difficult to
compare with our simple model.

\cite{Momany2006} studied the distribution of RGB stars to model
the outer Galaxy and attempt to explain the Canis Major
overdensity. They argue for no outer disc truncation, and that the
presence of a strong warp and flare at longitudes of about
240$^\circ$ explain well the overdensity. The study of this
particular sub-structure as well as the Monoceros ring is
postponed to a future paper.

Figure~\ref{warpwithR} compares the amplitude of the warp from
different authors with our result for three age class. A good
agreement is obtained with \cite{Burton1988}, \cite{LC2002} and
\cite{Yusifov2004} for stars with mean age equal to 2.5 Gyr in the
distance range, 10.5 $< \emph{R} <$ 13.5 kpc. However, the warp
slopes were determined using tracers in different Galactocentric
distance ranges. For instance, \cite{LC2002} and
\cite{Yusifov2004} measurements are claimed by their authors to be
valid up to 15 and 18 kpc, respectively. There is also a good
agreement between \cite{DS2001} and our results for mean age equal
to 6.0 Gyr for 9.0 $< R <$ 11.5 kpc. In addition, our estimates of
the warp for stars with mean age equal to 0.575 Gyr at negative
\emph{z$_{\rm w}$} are in good agreement with \cite{Levine2006},
at less than $1.5-\sigma$. The agreement is less good at positive
values of \emph{z$_{\rm w}$}, as if the warp effect on stars was
different from the HI warp.

We conclude that there is a general consistency between our
results and previous ones. The difference between our estimates of
the warp shape and other authors can be attributed to the
different ranges of longitudes and latitudes covered in each
study, as well as to the tracers and methods used. But we are
first to claim a warp dependence with the age of the tracers,
which is a clue for the scenario of warp formation.

\subsection{How does the Galaxy flare in the outskirts}

It has been long shown that the HI gas is flaring in the outer
Galaxy. This is most probably related to the vertical force
\citep{Creze1998,Holmberg2000,Kerr1986,Moni2012,Siebert2003,Sanchez2011},
which is dominated in the inner Galaxy by the stellar disc but in
the outer galaxy by dark matter. Hence a good characterization of
the \emph{K$_{\rm z}$} and the flare in the outer Galaxy
would lead to constrains on its dark matter content.

\cite{{Kalberla2008}} studied the HI distribution in the outer
Milky Way and showed that the gas is distributed in two
populations. The main gas layer goes to \emph{R} $\sim$ 35 kpc;
the second one goes up to 60 kpc with a very shallow scale length
of 7.5 kpc. The first and dominant component is flaring and
lopsided. They explain the asymmetry by a dark matter wake towards
the Magellanic Clouds. This is much shallower than the flare
indicated (see below) by \cite{LC2002}.

\cite{Alard2000} using 2MASS data studied the flare and warp for
longitudes located at $\ell \sim 66.0^\circ$, $\ell \sim
180.0^\circ$ and $\ell \sim 240.0^\circ$ for $|b| < 50^\circ$
founding evidences for an asymmetry associated with the Galactic
warp. He also argued that the flaring and warping seen in the
stellar disc is very similar to the characteristics observed in HI
disc. \cite{LC2002} estimated the flare distance scale to be 4.6
$\pm$ 0.5 kpc, which gives an increase of the scale height by a
factor of two at a Galactocentric distance of about 11 kpc and a
factor of 10 at 18 kpc. Instead of an exponential, we are using a
linear increase of the scale height with Galactocentric radius,
which gives a factor of two at 12.2 kpc close to \cite{LC2002},
and a factor of 4 at about 20 kpc, a less extreme value.

\begin{figure}[h]
\includegraphics[scale=0.4,angle=90]{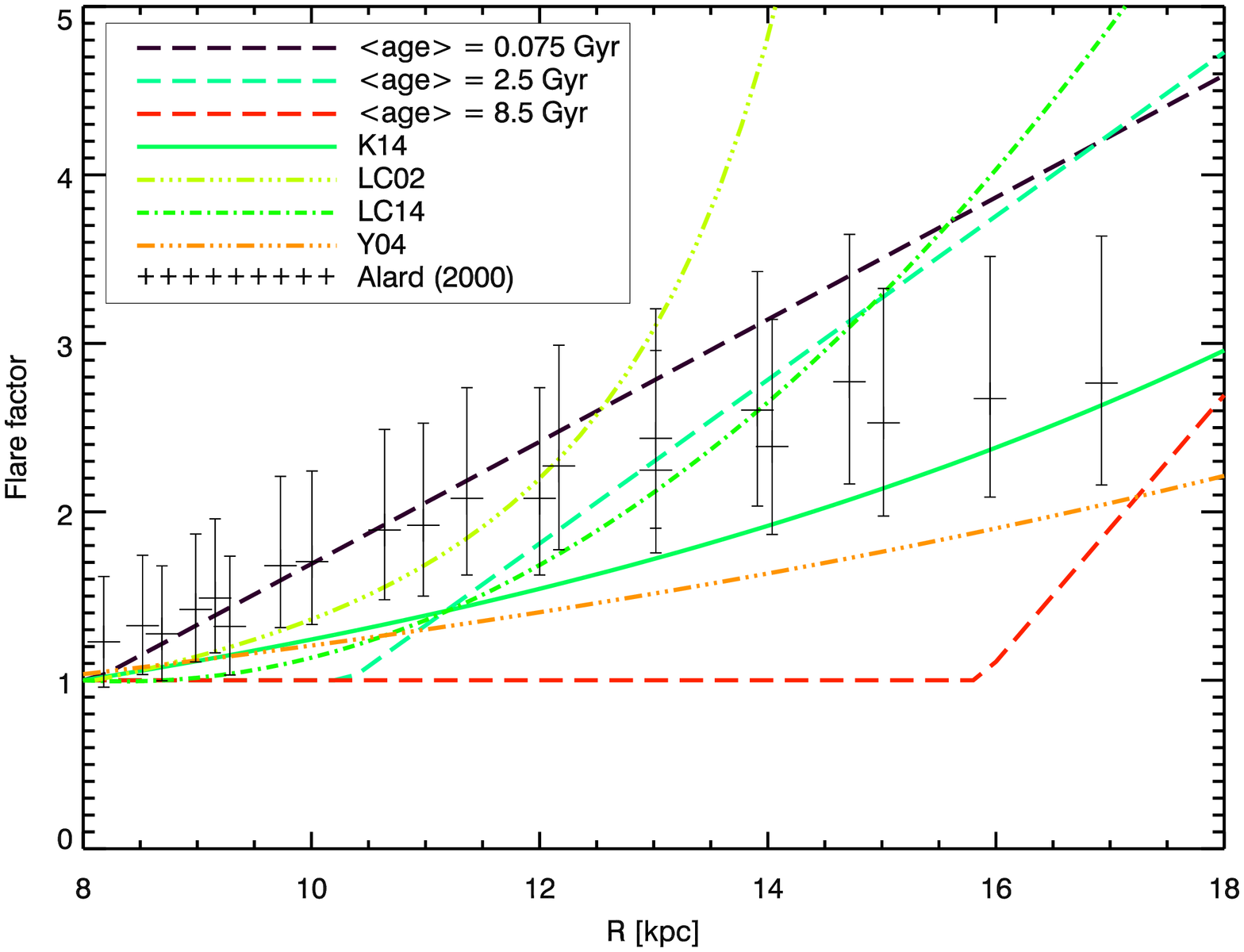}
\includegraphics[scale=0.4,angle=90]{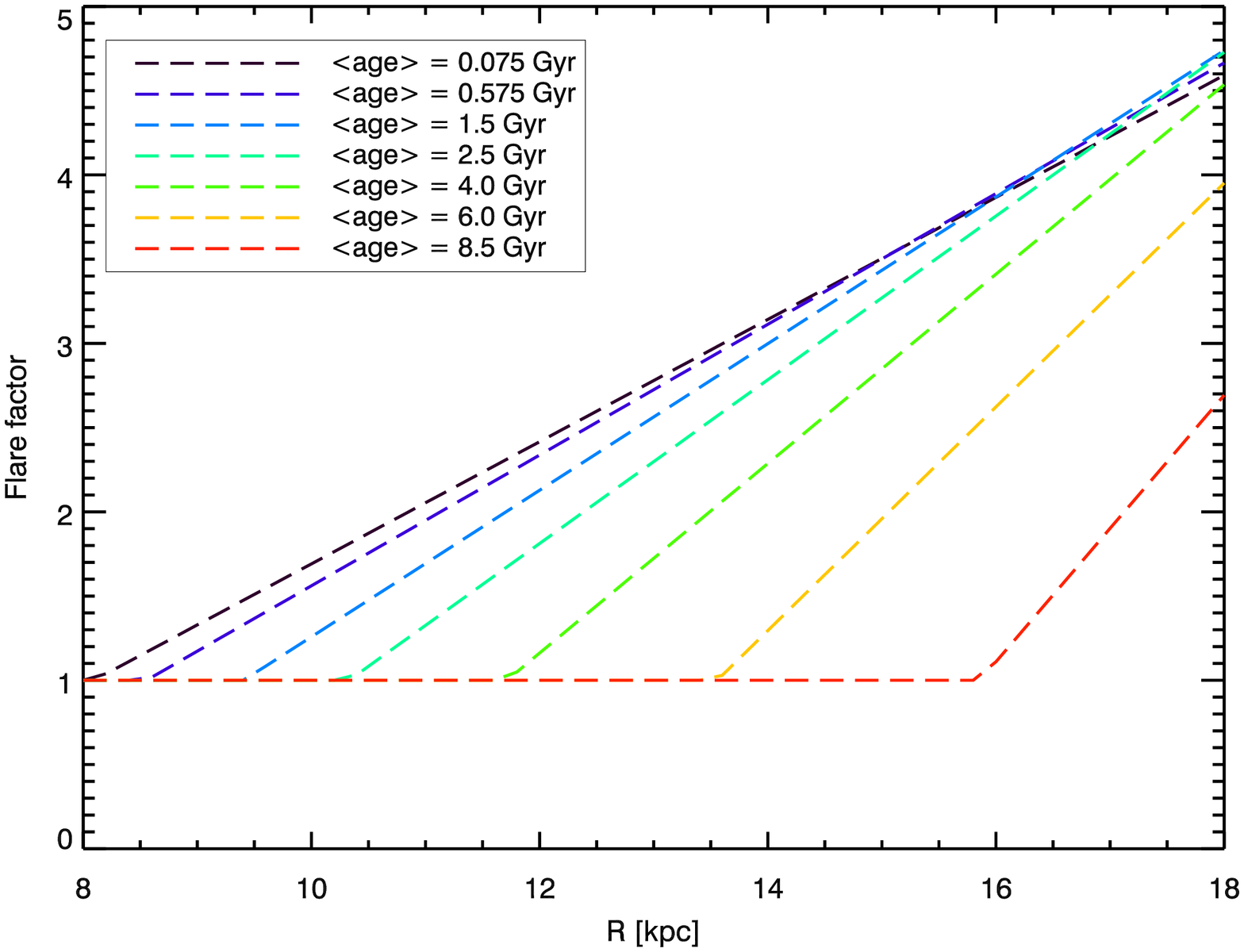}
\centering \caption{Upper panel: comparison of flare factors (see
text for its definition), lines: from \cite{LC2002} from red
giants (LC02), \cite{Yusifov2004} from pulsars (Y04),
\cite{Kalberla2014} from HI gas (K14), \cite{LC2014} from
SDSS-SEGUE data (LC14), and our study (dashed lines). Crosses are
the data of \cite{Alard2000} from 2MASS for three range of
longitudes: $\ell$ = 66.0$^\circ$, 180$^\circ$ and 240$^\circ$.
The values of $h_{\rm z}$(R$_{\rm Sun}$) considered for each author were
0.285, 0.580, 0.200, 0.240 and 0.250 pc, respectively. Lower
panel: Flare factor for each age class bin in our best fit model from
sim-2.} \label{flarewithR}
\end{figure}

Figure~\ref{flarewithR} (upper panel) shows a comparison of our
flare factor (Eqn 2) with the one of the \cite{Alard2000},
\cite{LC2002}, \cite{Yusifov2004}, \cite{Kalberla2014},
\cite{LC2014}. To compute the flare factor and compare it with our
results, we have divided the flare from other authors by h$_{\rm
z}$(R$_{\rm Sun})$ when the flare factor was not given. The values
of h$_{\rm z}$(R$_{\rm Sun})$ were taken from Eq. 4 of
\cite{Kalberla2014}, Eq. 3 of \cite{LC2014}, the Eqn quoted in the
abstract of \cite{LC2002} and Eq. 7 of \cite{Yusifov2004}. The
values of h$_{\rm z}$(R$_{\rm Sun}$) used by each author are
indicated in the caption of Figure~\ref{flarewithR}.

Concerning the estimations of errors on the flare factor,
\cite{Yusifov2004} mentioned an error around 30\% on the pulsar
distances determination, \cite{LC2002} estimated random errors
ranging from 5 to 7\%, and \cite{LC2014} estimated the error to be
approximately 18\%.

At small distances, all studies give very similar flare factors.
In young stars ($<\rm age>~=$ 0.075 Gyr) our flare factor appears
similar to \cite{LC2002} flare up to 13.0 kpc, and the values are
slightly larger for the older stars. For intermediate age stars
($<\rm age>~=$ 2.5 Gyr), a good agreement is seen for a wide range
of Galactocentric radius (10.5 kpc $\le$ \emph{R} $<$ 16.0 kpc) in
comparison with \cite{LC2014}. These authors determine the flare
from the SEGUE imaging survey. They fitted a model to F8-G5 stars
selected by colours in a sample which did not cover low latitudes
$|b|<$ 8$^\circ$, contrarily to our study. They claimed to have
found a significant flare but no outer disc truncation. Their
flare factor amounts to 3.7 for thin disc stars between the Sun
and a Galactocentric distance of 15 kpc, and a factor of 8.3 at a
distance of 20 kpc. In comparison, our result points to a factor
of 3.8 at 15 kpc and 5.8 at 20 kpc. We assumed a linear slope of
the flare to limit the number of parameters. The agreement up to
15 kpc is remarkable. The difference for further distance is most
probably due to the disc truncation that we see in-plane.

\cite{Yusifov2004} has a flare scale length of 14 kpc which gives
a shallow increase of the scale height with Galactocentric radius,
shallower than the HI flare. However, their sample is small (1600
pulsars) and probably incomplete in the outer Galaxy. No estimate
of the error bars was given. \cite{Hammersley2011} by using five
SDSS fields with $|b| > 11^\circ$ found a smaller amplitude for
the flare than obtained, for instance, by \cite{Yusifov2004}.

Recently, \cite{Feast2014} identified a few Cepheids at a large
distance from the Galactic plane. These Cepheids are a clear
example that the young disc is also flaring.

\cite{Michev2015} attempt to study the effect of the flare on the
radial gradient of age, that can be seen in the "geometrical"
thick disc (as defined by the population located at some distance
from the Galactic plane, where one expects to be dominated by
thick disc stars). They show from numerical simulations that a
radial gradient in age naturally emerges due to the flaring of the
populations, if younger populations are more extended than older
ones. \cite{Martig2016} by using APOGEE data reinforces the
results found by Minchev et al. (2015). In our study, the younger
disc populations are indeed more extended than older ones, and
they also flare earlier. Hence it is expected to produce such age
radial gradient at a few kiloparsec distances from the plane. We
have not tested this yet, but it will be considered in a future
study.

\section{Conclusions}

We have studied the outer Galaxy structure by using the
Besan\c{c}on Galaxy Model with 2MASS data in order to constrain
external disc parameters, such as warp and flare shape, scale
length and disc truncation parameters. After the parameter
optimisation, the relative difference between the model and 2MASS
is generally 10-20\% at most, which includes the cosmic variance
of the star counts as well as the effect of patchy extinction.

We show strong evidence that the thin disc scale length, as well
as the warp and flare shapes, changes with age, and proposed
expressions for these dependencies. These results impact directly
our comprehension about, not only the shape of those components,
but also their origin.

The warp can come from misalignment of the disc inside the dark
halo, which might change with time (precession). Alternatively, it
can be due to the interaction with the Magellanic Clouds. In both
cases, it can imply that tracers of different ages show different
spatial distribution, either because there is precession, or
because they react differently to the perturbation. The flare
could be caused by the fact that the intergalactic gas (or gas
coming from Galactic fountains) is falling into the disc more
slowly and on longer time scales in the outskirts than in the
inner Galaxy. We show in the present work some evidence for their
dependence with age that reinforces the point that the warp has a
dynamical origin as also demonstrated by \cite{LC2007}, among
others.

Our results clearly show the variation of scale length with the
age. The larger scale length for youngest stars (3.9 kpc) is well
in agreement with the values found in the HI gas, while the
shorter value (2 kpc) for the oldest thin disc is similar to the
one of the thick disc. This also directly affects our
comprehension of the history of Galactic formation and evolution,
reinforcing the idea of an inside-out process of formation.

We also found a disc truncation with \emph{R$_\mathrm{dis}$} =
16.1 $\pm$ 1.3 kpc. But when we allow the parameters to vary with
time, the disc truncation is no more clear and could be at larger
distances than 19 kpc, a distance where the stars are scarce
anyway, simply due to the exponential fall off. However, this
value is not completely independent of other parameters such as
the flare and disc scale length.

A further study involving higher latitudes will be considered to
strengthen the conclusion about the disc truncation and its
dependence with age.

\begin{acknowledgements}
We thank Misha Haywood, Paola Di Matteo and Vanessa Hill for
fruitful discussions about this work. We also thank the anonymous
referee for useful, valuable, and detailed comments on the
manuscript. Eduardo Am\^{o}res would like to thank very much all
staff of the Observatoire de Besan\c{c}on for the hospitality and
courtesy during his sejour and collaboration visits. We also
thanks Dr Douglas Marshall for let available the 3D extinction
model prior to its publication. BGM simulations were executed on
computers from the UTINAM Institute of the Universit\'e Bourgogne
Franche-Comt\'{e}, supported by the R\'egion de Franche-Comt\'e
and Institut des Sciences de l'Univers (INSU). Financial support
for this work was provided by Conselho Nacional para o
Desenvolvimento Cient\'ifico e Tecnol\'ogico (CNPq), the
Coordenac\~ao de Aperfei\c{c}oamento Pessoal de N\'ivel Superior
(CAPES) and the CNRS. Eduardo Brescansin de Am\^{o}res had a CNPq
fellowship (2003/201113-3) and also thanks INCT$-$A (Brazil). This
publication made use of data products from the Two Micron All Sky
Survey, which is a joint project of the University of
Massachusetts and the Infrared Processing and Analysis
Center/California Institute of Technology, funded by the National
Aeronautics and Space Administration and the National Science
Foundation.
\end{acknowledgements}

\bibliographystyle{aa} 
\bibliography{EBAmoresetal_Twocolumns_Rf_LE.bbl} 

\bibliographystyle{aa}   
\begin{thebibliography}{99}
\bibitem[{{Abedi} {et~al.}(2014)}]{Abedi2014} {{Abedi}, H. and {Mateu}, C. and {Aguilar}, L.~A. and {Figueras}, F. and
    {Romero-G{\'o}mez}, M.}, 2014, \mnras, 442, 3627
\bibitem[{{Alard}(2000)}]{Alard2000}{{Alard}, C.}, 2000, {astro-ph/0007013}
\bibitem[{{Am{\^o}res} \& {L{\'e}pine}(2005)}]{AL2005}
{Am{\^o}res}, E.~B., {L{\'e}pine}, J.~R.~D., 2005, \aj, 130,
659
\bibitem[{{Bertin}(1996)}]{Bertin1996}{{Bertin}, E}, 1996, PhD Thesis, Universit\'e Paris VI
\bibitem[{{Bilir} {et~al.}(2006)}]{Bilir2006}{{Bilir}, S.,{Karaali}, S., {Ak}, S., {Yaz}, E.,
{Hamzao{\u g}lu}, E.}, 2006, \na, 12, 234
\bibitem[{{Bienaym\'{e}} {et~al.}(1987)}]{Bienayme1987a}{{Bienaym\'e}, O.,{Robin}, A.C.,
\& {Cr\'ez\'e}, M.}, 1987, \aap, 180, 94
\bibitem[{{Bovy} {et~al.}(2012)}]{bovy2012} {{Bovy}, J., {Rix}, H-W, {Liu}, C., {Hogg}, D. W.,
{Beers},T., {Lee}}, Y. S., 2012, \apj, 753, 148
\bibitem[{{Bovy \& Rix}(2013)}]{Bovy2013}{{Bovy}, J., R., {Hans-Walter}}, 2013, \apj, 779, 115
\bibitem[{{Burton \& te Lintel Hekkert}(1986)}]{Burton1986}{{Burton}, W.~B. \& {te Lintel Hekkert}, P.}, 1986, {\aaps}, 65, 427
\bibitem[{{Burton}(1988)}]{Burton1988}{{Burton}, W. B.}, 1988, in Galactic and Extragalactic Radio Astronomy,
2nd version, p. 295, Springer-Verlag
\bibitem[{{Brook} {et~al.}(2012)}]{Brook2012}{{Brook}, C.~B., {Stinson}, G.~S., {Gibson}, B.~K.,
    {Kawata}, D., {House}, E.~L., {Miranda}, M.~S., {Macci{\`o}}, A.~V.,
    {Pilkington}, K., {Ro{\v s}kar}, R., {Wadsley}, J.,
    {Quinn}, T.~R.}, 2012, \mnras, 426, 690
\bibitem[{{Cabrera-Lavers} {et~al.}(2005)}]{Cabrera-Lavers2005}{{Cabrera-Lavers}, A., {Garz\'{o}n}, F., {Hammersley}, P.
L.}, 2005, \aap, 433, 173
\bibitem[{{Cambr\'{e}sy} {et~al.}(2003)}]{Cambresy2003}{{Cambr\'{e}sy}, L., {Beichman}, C.A., {Jarrett}, T.H., {Cutri}, R.M.}, 2003, \aj, 123, 2559
\bibitem[{{Chang} {et~al.}(2011)}]{Chang2011}{{Chang}, C., {Ko}, C., {Peng}}, T., 2011, \apj, 740, 34
\bibitem[{{Charbonneau}(1995)}]{Charb1995}{{Charbonneau}, P.}, 1995, {\apjs}, 101, 309
\bibitem[{{Charbonneau} \& {Knapp}(1996)}]{Charbonneau1996}{{Charbonneau}, P., {Knapp}, B.}, 1996, A Users Guide to PIKAIA 1.0: NCAR Technical Note
418+IA
\bibitem[{{Chen} {et~al.}(2001)}]{Chen2001}{{Chen}, B., {SDSS Collaboration}}, 2001, \apj, 553, 184
\bibitem[{{Cheng} {et~al.}(2012)}] {Cheng2012} {{Cheng}, J.~Y., {Rockosi}, C.~M., {Morrison}, H.~L.,
{Lee}, Y.~S., {Beers}, T.~C., {Bizyaev}, D., {Harding}, P.,
{Malanushenko}, E., {Malanushenko}, V., {Oravetz}, D., {Pan}, K.,
{Schlesinger}, K.~J., {Schneider}, D.~P., {Simmons}, A., {Weaver},
B.~A.}, 2012, \apj, 752, 51
\bibitem[{{Cignoni} {et~al.}(2008)}]{Cignoni2008} Cignoni, M., Tosi, M., Bragaglia, A., Kalirai, J. S., Davis, D.
S., 2008, \mnras, 386, 2235
\bibitem[{{Czekaj} {et~al.}(2014)}]{Czekaj2014}{{Czekaj}, M.~A., {Robin}, A.~C., {Figueras}, F., {Luri}, X.,
    {Haywood}, M.}, 2014, \aap, 564, A102
\bibitem[{{Cr\'{e}z\'{e}} et al. (1998)}]{Creze1998} {{Cr\'{e}z\'{e}}, M., {Chereul}, E., {Bienaym\'{e}}, O., {Pichon}, C.},
1998, {\aap}, 329, 920
\bibitem[{{Dehnen \& Binney}(1998)}]{Dehnen1998}{{Dehnen}, W., {Binney}, J.}, 1998, \mnras, 294, 429
\bibitem[Derri\`ere \& Robin(2001)]{Derriere2001} Derri\`ere, S.~\&
Robin, A.~C., ASP Conf.~Ser.~232: The New Era of Wide Field
Astronomy, p. 229, 2001
\bibitem[{{Diplas \& Savage}(1991)}]{Diplas1991}{{Diplas}, A., {Savage}, B.D.}, 1991, \apj, 377, 126
\bibitem[{{Drimmel}(2000)}]{Drimmel2000}{{Drimmel}, R.}, 2000, \aap, 358, L13
\bibitem[{{Drimmel \& Spergel}(2001)}]{DS2001}{{Drimmel}, R., {Spergel}, D.~N.}, 2001, \apj, 556, 181
\bibitem[{{Feast} {et~al.}(2014)}]{Feast2014}{{Feast}, M.~W., {Menzies}, J.~W., {Matsunaga}, N.,
    {Whitelock}, P.~A.}, 2014, {\nat}, 509, 342
\bibitem[{{Gyuk} {et~al.}(1999)}]{Gyuk1999}{{Gyuk}, G., {Flynn}, C., {Evans}, N.~W.}, 1999, \apj, 521, 190
\bibitem[{{Grabelsky} {et~al.}(1987)}]{Grabelsky1987}{{Grabelsky}, D.~A., {Cohen}, R.~S., {Bronfman}, L.,
    {Thaddeus}, P., {May}, J.}, \apj, 315, 122
\bibitem[{{G{\'o}mez} {et~al.}(1997){G{\'o}mez}, {Grenier}, {Udry}, {Haywood},
  {Meillon}, {Sabas}, {Sellier}, \& {Morin}}]{Gomez}
{G{\'o}mez}, A.~E., {Grenier}, S., {Udry}, S., {et~al.} 1997, in ESA Special
  Publication, Vol. 402, Hipparcos - Venice '97, ed. {R.~M.~Bonnet, E.~H{\o}g,
  P.~L.~Bernacca, L.~Emiliani, A.~Blaauw, C.~Turon, J.~Kovalevsky,
  L.~Lindegren, H.~Hassan, M.~Bouffard, B.~Strim, D.~Heger, M.~A.~C.~Perryman,
  \& L.~Woltjer}, 621--624
\bibitem[{{Freudenreich}(1998)}]{Freud1998}{{Freudenreich}, H. T.}, 1998, \apj, 492, 495
\bibitem[{{Habing}(1988)}]{Habing1988}{{Habing}, H.~J.}, 1988, \aap, 200,40
\bibitem[{{Haywood} {et~al.}(1997)}]{Haywood1997}{{Haywood}, M.,{Robin}, A.C.,{{Cr\'ez\'e}}, M}. 1997, \aap, 320, 440
\bibitem[{{Haywood} {et~al.}(2013)}]{Haywood2013}{{Haywood}, M., {Di Matteo}, P., {Lehnert}, M. D., {Katz}, D.,
{G\'{o}mez}, A.}, 2013, \aap, 560, A109
\bibitem[{{Hammersley \& L\'{o}pez-Corredoira}(2011)}]{Hammersley2011}{{Hammersley}, P.~L., {L{\'o}pez-Corredoira}, M.}, 2011, \aap, 527, A6
\bibitem[{{Henderson} {et~al.}(1982)}]{Henderson1982}{{Henderson}, A.P., {Jackson}, P.D., {Kerr}, F.J.}, 1982, \apj, 263, 116
\bibitem[{{Holand}(1975)}]{Holand1975}{{Holand}, J.~H.},1975, Adaptation in Natural and Artificial Systems (Ann Arbor: Univ. Michigan
Press)
\bibitem[{{Holmberg} \& {Flynn}(2000)}]{Holmberg2000}{{Holmberg}, J., {Flynn}, C.}, 2000, \mnras, 313, 209
\bibitem[{{Juri\'{c}} {et~al.}(2008)}]{Juric2008}{{Juri\'{c}}, M.,{et~al.}}, 2008, \apj, 673, 864
\bibitem[{{Kennicutt}(1989)}]{Kennicutt1989}{{Kennicutt}, R. C. Jr.}, 1989, {\apj}, 344, 685
\bibitem[{{Kalberla} {et~al.}(2005)}]{Kalberla2005}{{Kalberla}, P.~M.~W., {Burton}, W.~B., {Hartmann}, D.,
    {Arnal}, E.~M., {Bajaja}, E., {Morras}, R., {P{\"o}ppel}, W.~G.~L.}, 2005, \aap, 440, 775
\bibitem[{{Kalberla} \& {Dedes}(2008)}]{Kalberla2008}{{Kalberla, P. M. W., {Dedes},
L.}, 2008}, \aap, 487, 951
\bibitem[{{Kalberla} \& {Kerp}(2009)}]{Kalberla2009}{{Kalberla}, P.~M.~W., {Kerp}, J.}, 2009, {\araa}, 47, 27
\bibitem[{{Kalberla} {et~al.}(2014)}]{Kalberla2014}{{Kalberla}, P.~M.~W., {Kerp}, J., {Dedes}, L., {Haud}, U.}, 2014, {\apj}, 794, 90
\bibitem[{{Karaali} {et~al.}(2007)}]{Karaali2007}{{Karaali}, S., {Bilir}, S., {Yaz}, E., {Hamzao{\u g}lu}, E.,
    {Buser}, R.}, 2007, {\pasa}, 24, 208
\bibitem[{{Kerr} \& {Lynden-Bell}}(1986)]{Kerr1986}{{Kerr}, F.~J., {Lynden-Bell}, D.}, 1986, {\mnras}, 221, 1023
\bibitem[{{Larsen} \& {Humphreys}(2003)}]{Larsen2003}{{Larsen}, J.~A., {Humphreys}, R.~M.}, 2003, \aj, 125, 1958
\bibitem[{{Larson}(1976)}]{Larson1976}{{Larson, R. B.}}, 1976, \mnras, 176, 31
\bibitem[{{Levine} {et~al.}(2006)}]{Levine2006}{{Levine}, E. S., {Blitz}, L., {Heiles}, C.}, 2006, \apj, 643, 881
\bibitem[{{L\'{o}pez-Corredoira} {et~al.}(2002)}]{LC2002}{{L{\'o}pez-Corredoira}, M., {Cabrera-Lavers},
A., {Garz{\'o}n}, F., {Hammersley}, P.~L.}, 2002, \aap, 394, 883
\bibitem[{{L\'{o}pez-Corredoira} {et~al.}(2007)}]{LC2007}{{L\'{o}pez-Corredoira}, M.,
 {Momany}, Y., {Zaggia}, S., {Cabrera-Lavers}, A.}, 2007, \aap, 472, L47
\bibitem[{{{L{\'o}pez-Corredoira} \& {Molg{\'o}}}(2014)}]{LC2014}{{L{\'o}pez-Corredoira}, M.,  {Molg{\'o}}, J.},, 2014, \aap, 567, A106
\bibitem[{{Manchester} {et~al.}(2005)}]{Manchester2005}{{Manchester}, R.~N., {Hobbs}, G.~B., {Teoh},
A., {Hobbs}, M.}, 2005, \aj, 129, 1993
\bibitem[{{Marshall} {et~al.}(2006)}]{Marshall2006}{{Marshall}, D. J., {Robin}, A. C., {Reyl\'{e}}, C.,
{Schultheis}, M., {Picaud},S.}, 2006, \aap, 435, 635
\bibitem[{{Marshall}(2009)}]{Marshall2009}{{Marshall}, D. J.}, 2009, private communication
\bibitem[{{Martig} {et~al.}(2016)}]{Martig2016}{{{Martig}, M. and {Minchev}, I. and {Ness}, M. and {Fouesneau}, M. and
    {Rix}, H.-W.}}, {arXiv:1609.01168}, 2016
\bibitem[{{Martin} {et~al.}(2004)}]{Martin2004}{{Martin}, N. {et~al.}}, \mnras, 348, 12, 2004
\bibitem[{{May} {et~al.}(1997)}]{May1997}{{May}, J., {Alvarez}, H., {Bronfman}, L.}, 1997, \aap, 327, 325
\bibitem[{{McMillan}(2011)}]{McMillan2011}{{McMillan}, P. J.}, 2011, \mnras, 414, 2446
\bibitem[{{Mitchell}(1996)}]{Mitchell1996}{{Mitchell}, M., 1996}, An Introduction to Genetic Algorithms (Cambridge, MIT Press)
\bibitem[{{Minchev} {et~al.}(2015)}]{Michev2015}{{Minchev}, I. and {Martig}, M. and {Streich}, D. and {Scannapieco}, C. and 
    {de Jong}, R.~S. and {Steinmetz}, M.}, 2015, \apjl, 804, L9
\bibitem[{{Momany} {et~al.}(2006)}]{Momany2006}{{Momany}, Y., {Zaggia}, S., {Gilmore}, G., {Piotto}, G.,
    {Carraro}, G., {Bedin}, L.~R., {de Angeli}, F.}, 2006, \aap, 451, 515
\bibitem[{{Moni Bidin} {et~al.}(2012)}]{Moni2012}{{Moni Bidin}, C., {Carraro}, G., {M{\'e}ndez}, R.~A., {Smith}, R.}, 2012, {\apj}, 751
\bibitem[{{Nakanishi \& Sofue}(2003)}]{Nakanishi2003}{{Nakanishi}, H., {Sofue}, Y.} 2003, \pasj, 55, 191
\bibitem[{{Newberg} {et~al.}(2002)}]{Newberg2002} {{Newberg}, H.~J., {Yanny}, B., {Rockosi}, C., {Grebel}, E.~K.,
{Rix}, H.-W., {Brinkmann}, J., {Csabai}, I., {Hennessy}, G., {Hindsley}, R.~B., {Ibata}, R., {Ivezi{\'c}}, Z., {Lamb}, D.,
{Nash}, E.~T., {Odenkirchen}, M., {Rave}, H.~A., {Schneider}, D.~P., {Smith}, J.~A., {Stolte}, A., {York}, D.~G.}, 2002, \apj,
569, 245
\bibitem[{{Ng}(1998)}]{Ng1998}{{Ng}, Y.~K.}, 1998, \aaps, 132, 133
\bibitem[{{Ng} {et~al.}(2002)}]{Ng2002}{{Ng}, Y.~K., {Brogt}, E., {Chiosi}, C. {Bertelli}, G.}, 2002, \aap, 392, 1129
\bibitem[{{Ojha}(2001)}]{Ojha2001}{{Ojha}, D. K.}, 2001, \mnras, 322, 426
\bibitem[{{Perryman} {et~al.}(2014)}]{Perryman2014}{{Perryman}, M., {Spergel}, D.~N., {Lindegren}, L.}, 2014, \apj, 789, 166
\bibitem[{{Porcel}{et~al.}(1998)}]{Porcel1998}{{Porcel}, C., {Garz{\'o}n}, F., {Jimenez-Vicente}, J., {Battaner}, E.}, 1998,
\aap, 330, 136
\bibitem[{{Rahimi} {et~al.}(2011)}]{Rahimi2011}{{Rahimi}, A., {Kawata}, D., {Allende Prieto}, C. {et~al.}}, 2011, \mnras,
415, 1469
\bibitem[{{Reshetnikov} {et~al.}(2016)}]{Reshetnikov2016}{{Reshetnikov}, V.~P. and {Mosenkov}, A.~V. and {Moiseev}, A.~V. and
    {Kotov}, S.~S. and {Savchenko}, S.~S.}, 2016, \mnras, 461, 4233
\bibitem[{{Reyl\'{e}} {et~al.}(2009)}]{Reyle2009}{{Reyl{\'e}}, C., {Marshall}, D.~J., {Robin}, A.~C.,
    {Schultheis}, M.}, 2009,\aap, 2009, 495, 819
\bibitem[{{Robin} \& {Cr\'{e}z\'{e}}(1986)}]{Robin1986}{{Robin},~A.~C., {Cr\'{e}z\'{e}}, M.}, 1986, \aap, 157, 71
\bibitem[{{Robin} {et~al.}(1992a)}]{Robin1992a}{{Robin},~A.~C., {Cr\'{e}z\'{e}} M., {Mohan}, V.}, 1992, \aap, 265, 32
\bibitem[{{Robin} {et~al.}(1992b)}]{Robin1992b}{{Robin},~A.~C., {Cr\'{e}z\'{e}}, M., {Mohan}, V.}, 1992, \apjl, 400, L25
\bibitem[{{Robin} {et~al.}(2003)}]{Robin2003}{{Robin},~A.~C., {Reyl\'{e}},C.,{Derri{\`e}re},S.,{Picaud}, S.}, 2003, \aap, 409, 523
\bibitem[{{Robin} {et~al.}(2012)}]{Robin2012}{{Robin},~A.~C., {Marshall}, D.~J., {Schultheis}, M.,
{Reyl{\'e}}, C.}, 2012, {\aap}, 538, A106
\bibitem[{{Robin}~{et~al.}(2014)}]{Robin2014}{{Robin},~A.~C., {Reyl{\'e}}, C., {Fliri}, J., {Czekaj}, M.,
    {Robert}, C.~P. and {Martins}, A.~M.~M.}, 2014, \aap , 569, {A13}
\bibitem[{{Rocha-Pinto} {et~al.}(2003)}]{Rocha-Pinto2003}{{Rocha-Pinto}, H.~J., {Majewski}, S.~R.,
{Skrutskie}, M.~F., {Crane}, J.~D.}, 2003, \apj, 594, L115
\bibitem[{{Ro{\v s}kar} {et~al.}(2010)}]{Roskar2010}{{Ro{\v s}kar}, R. {Debattista}, V.~P. {Brooks}, A.~M.
    {Quinn}, T.~R. {Brook}, C.~B. {Governato}, F. {Dalcanton}, J.~J.
    {Wadsley}, J.}, 2010, \mnras, 408, 783
\bibitem[{{Ruphy} {et~al.}(1996)}]{Ruphy1996}{{Ruphy}, S., {Robin}, A.~C., {Eptchein}, N. {et~al.}}, 1996, \aap, 313, L21
\bibitem[{{Sale} {et~al.}(2010)}]{Sale2010}{{Sale}, S. E., et al.}, 2010, \mnras, 402, 713
\bibitem[{{Schwarz} {et~al.}(1978)}]{Schwarz1978}{{Schwarz}, G. E.}, 1978, Annals of Statistics, 6, 461
\bibitem[{{Seiden} {et~al.}(1984)}]{Seiden1984}{{Seiden}, P.~E., {Schulman}, L.~S., {Elmegreen}, B.~G.}, 1984, \apj, 282, 95
\bibitem[{{S{\'a}nchez-Salcedo} {et~al.}(2011)}]{Sanchez2011} {{S{\'a}nchez-Salcedo}, F.~J., {Flynn},
C., {Hidalgo-G{\'a}mez}, A.~M.}, 2011, {\apjl}, 731, L35
\bibitem[{{Sevenster} {et~al.}(1999)}]{Sevenster1999}{{Sevenster}, M., {Saha}, P., {Valls-Gabaud}, D., {Fux}, R.}, 1999, \mnras, 307, 584
\bibitem[{{Shen} \& {Sellwood}(2006)}]{Shen2006}{{Shen}, J., {Sellwood}, J.~A.}, 2006, \mnras, 370, 2
\bibitem[{{Siebert} {et~al.}(2003)}]{Siebert2003}{{Siebert}, A., {Bienaym{\'e}}, O., {Soubiran}, C.}, 2003, {\aap}, 399, 531
\bibitem[{{Siegel} {et~al.}(2002)}]{Siegel2002}{{Siegel}, M.~H., {Majewski}, S.~R., {Reid}, I.~N.,
    {Thompson}, I.~B.}, 2002, \apj, 578, 151
\bibitem[{{Skrutskie} {et~al.}(2006)}]{Skrutskie2006}{{Skrutskie}, M.~F., {et~al.}}, 2006, \aj, 131, 1163
\bibitem[{{Sommer-Larsen} {et~al.}(2003)}]{Sommer2003}{{Sommer-Larsen}, J., {Gotz}, M., {Portinari}, L.}, 2003, \apj, 596, 47
\bibitem[{{van der Kruit}(1979)}]{vanderkruit1979}{{van der Kruit}, P.~C.}, 1979, \aaps, 38, 15
\bibitem[{{Weinberg \& Blitz}(2006)}]{Weinberg2006}{{Weinberg}, M. D., {Blitz},
L.}, 2006, \apj, 641, L33
\bibitem[{{Wouterloot} {et~al.}(1990)}]{Wouterloot1990}{{Wouterloot}, J.~G.~A., {Brand}, J., {Burton}, W.~B.,
    {Kwee}, K.~K.}, 1990, \aap, 230, 21
\bibitem[{{Xu} {et~al.}(2015)}]{Xu2015}{{Xu}, Y., {Newberg}, H.~J., {Carlin}, J.~L., {Liu}, C.,
    {Deng}, L., {Li}, J., {Sch{\"o}nrich}, R., {Yanny}, B.}, 2015, \apj, 801, 105
\bibitem[{{Yaz} \& {Karaali}(2010)}]{Yaz2010}{{Yaz}, E., {Karaali}, S.}, 2010, \na, 15, 234
\bibitem[{{Yusifov}(2004)}]{Yusifov2004} Yusifov I., 2004, in Uyaniker B., Reich W., Wielebinski R.,
eds, The Magnetized Interstellar Medium. Copernicus GmbH, Katlenburg-Lindau, p. 165
\end{thebibliography}

\clearpage
\newpage

\appendix

\section{Parameters correlations}

\begin{figure}[ht]
\centering
\includegraphics[scale=0.75,angle=90]{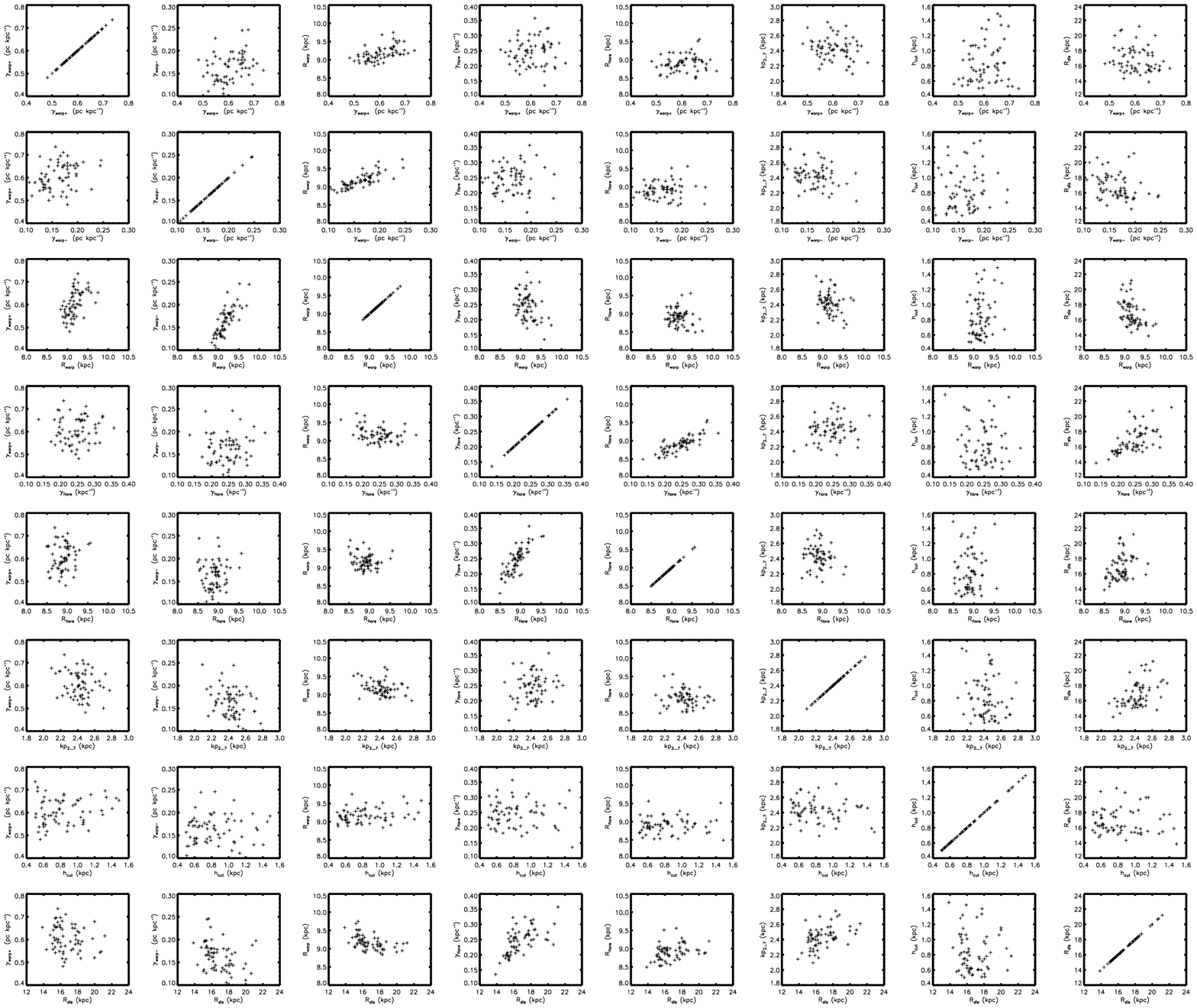}
\caption{Correlations between parameters for sim-1 for 100
independent runs.} \label{AppendixSim1}
\end{figure}

\newpage
\clearpage

\end{document}